\documentclass[11pt]{article}

\usepackage[T1]{fontenc}
\usepackage{polyglossia}
\usepackage[normalem]{ulem}
\usepackage{amsmath}
\usepackage{amssymb}
\usepackage{amsthm}
\usepackage{mathtools}
\usepackage{mathabx}
\usepackage{slashed}
\usepackage{caption}
\usepackage{makecell}
\usepackage[colorlinks=false,
linkcolor=darkblue,  
urlcolor=blue,    
filecolor=blue,     
citecolor=red,
linktocpage=true,
pdfstartview=FitV,
bookmarksopen=true,
pdfencoding=auto,
pagebackref]{hyperref}
\usepackage{comment}
\usepackage{cite}
\usepackage{simpler-wick}
\usepackage{soul}
\usepackage[compat=1.1.0]{tikz-feynman}
\usepackage[a4paper, total={6in, 8in}]{geometry}
\usepackage{tikz}
\usetikzlibrary{decorations.markings}
\usetikzlibrary{shapes.misc}
\tikzset{cross/.style={cross out, draw=black, minimum size=2*(#1-\pgflinewidth), inner sep=0pt, outer sep=0pt},cross/.default={3pt}}

\makeatletter
\renewcommand\section{\@startsection {section}{1}{\z@}%
                                   {-3.5ex \@plus -1ex \@minus -.2ex}%
                                   {2.3ex \@plus.2ex}%
                                   {\normalfont\large\bfseries}}
\renewcommand\subsection{\@startsection{subsection}{2}{\z@}%
                                   {-3.25ex\@plus -1ex \@minus -.2ex}%
                                   {1.5ex \@plus .2ex}%
                                   {\normalfont\normalsize\bfseries}}

\makeatother
\numberwithin{equation}{section}

\definecolor{hyperref}{RGB}{026,028,185}

\definecolor{cardinal}{rgb}{0.6,0,0}
\definecolor{darkgreen}{rgb}{0,0.5,0}
\definecolor{golden}{rgb}{0.92, 0.7, 0}
\definecolor{midnight}{rgb}{0, 0, 0.5}
\definecolor{darkblue}{rgb}{0.2, 0, 0.8}

\begin{document}

\newcommand{\dd}{\mathrm{d}}
\newcommand{\del}{\partial}
\newcommand{\eps}{\varepsilon}
\newcommand{\Th}{\theta}
\newcommand{\tr}{\operatorname{tr}}
\newcommand{\Tr}{\operatorname{tr}}
\newcommand{\AdS}{\mathrm{AdS}_{d+1}}
\newcommand{\AdSd}{$\mathrm{AdS}_2$\ }
\newcommand{\dS}{$\mathrm{dS}_{d+1}$\ }
\newcommand{\dSD}{\mathrm{dS}_D}
\newcommand{\Sd}{\mathrm{S}^d}
\newcommand{\Sn}{\mathrm{S}^n}
\newcommand{\Sf}{\mathrm{S}^5}
\newcommand{\AdSfSf}{\mathrm{AdS}_5 \times \mathrm{S}^5}
\newcommand{\abs}[1]{\left\vert #1 \right\vert}
\newcommand{\sgnabs}[1]{\left\lfloor #1 \right\rceil}
\newcommand{\C}{\mathbb{C}}
\newcommand{\R}{\mathbb{R}}
\newcommand{\Z}{\mathbb{Z}}
\newcommand{\N}{\mathbb{N}}
\newcommand{\OO}{\mathcal{O}}
\newcommand{\ket}[1]{| #1 \rangle}
\newcommand{\bra}[1]{\langle #1 |}
\newcommand{\braket}[2]{\langle #1 | #2 \rangle}
\newcommand{\nptf}[1]{\left\langle #1 \right\rangle}
\newcommand{\ch}[1]{\mathrm{ch}\left( #1 \right)}
\newcommand{\sh}[1]{\mathrm{sh}\left( #1 \right)}
\newcommand{\tah}[1]{\mathrm{th}\left( #1 \right)}
\newcommand{\sgn}[1]{\mathrm{sgn}\left( #1 \right)}
\newcommand{\unih}{\mathfrak{h}}
\newcommand{\unie}{\mathfrak{e}}
\newcommand{\unig}{\mathfrak{g}}
\newcommand{\flati}[1]{{\underline{#1}}}
\newcommand{\geleven}{\Gamma^{(11)}}

\newcommand{\be}{\begin{equation}}
\newcommand{\ee}{\end{equation}}

\vspace{ -3cm} \thispagestyle{empty} \vspace{-1cm}
\begin{flushright} 
\footnotesize
{HU-EP-26/20-RTG}
\end{flushright}%

\begin{center}
\vspace{1.2cm}
{\Large \bf \mathversion{bold}{Bremsstrahlung function  from Goldstone bosons in AdS$_2$}
}

\author{ABC\thanks{XYZ} \and DEF\thanks{UVW} \and GHI\thanks{XYZ}}
 \vspace{0.8cm} {
Federico Castellani$^{a}$ \footnote{\tt federico.castellani@physik.hu-berlin.de}, Lorenzo~Di Pietro$^{b,c}$ \footnote{\tt ldipietro@units.it}, Valentina~Forini$^{a}$ \footnote{\tt valentina.forini@hu-berlin.de}, Roman~Stemplowski$^{a}$ \footnote{{\tt roman.stemplowski@physik.hu-berlin.de}}}
 \vskip  0.5cm

\small
{\em
$^{a}$   
Institut f\"ur Physik, Humboldt-Universit\"at zu Berlin and IRIS Adlershof, \\
Zum Gro\ss en Windkanal 2, 12489 Berlin, Germany\\    
\vskip 0.02cm
$^{b}$   
Dipartimento di Fisica, Università di Trieste,
Strada Costiera 11, I-34151 Trieste, Italy   
\vskip 0.02cm
}
$^{c}$
{\em INFN, Sezione di Trieste, Via Valerio 2, I-34127 Trieste, Italy}
\normalsize
\end{center}

\vspace{0.3cm}
\begin{abstract} 
\noindent
We study quantum field theory on the \AdSd worldsheet of the fundamental string dual to the half-BPS Wilson line of $\mathcal{N}=4$ super Yang-Mills. We consider a recently proposed gauge-fixed string fluctuation Lagrangian around the AdS$_2$ minimal surface, and focus on the nonlinearly realized $SO(6)/SO(5)$ symmetry. The boundary two-point function of the associated tilt operators, whose normalization coefficient is proportional the Bremsstrahlung function, can be related to the bulk two-point function of the broken-symmetry Noether current through the boundary operator expansion.
We evaluate the one-loop diagrams contributing to the current correlator, discussing the role of dimensional regularization and recovering the known strong-coupling correction to the Bremsstrahlung function directly from AdS$_2$ worldsheet quantum field theory.

\end{abstract}
\overfullrule=0pt
\parskip=2pt
\parindent=12pt
\headheight=0in \headsep=0in \topmargin=0in \oddsidemargin=0in

\vspace{ -3cm} \thispagestyle{empty} \vspace{-1cm}

\newpage
\tableofcontents
\newpage

\section{Introduction and discussion} 

 A rigid Anti-de Sitter (AdS) background provides a useful way of organizing asymptotic data in Quantum Field Theory (QFT)~\cite{Polchinski:1999ry,Giddings:1999qu,Paulos:2016fap,Carmi:2018qzm}. The natural objects of study for QFT in AdS are boundary $n$-point functions, obtained from suitable limits of bulk Green functions and governed by the kinematics of a conformal theory in one lower dimension. 
 
 A particularly interesting asymptotic limit applies to conserved-currents operators in the case of spontaneous symmetry breaking (SSB). SSB in AdS is realized whenever a boundary condition at the conformal boundary of AdS is not invariant under a bulk global symmetry. When a symmetry is broken in this sense, the normal component of the corresponding current has a distinguished boundary limit, defining a boundary operator with protected scaling dimension. In the special case of a conformal field theory (CFT) in AdS, this is a familiar structure arising in the study of boundaries or defect CFT (DCFT). A conformal defect breaks part of the ambient symmetries, giving rise to protected defect operators, such as the displacement operator~\cite{Billo:2016cpy} (associated to the breaking of the bulk conformal symmetry) and tilt operators~\cite{Bray:1977tk, Herzog:2017xha, Cuomo:2021cnb, Padayasi:2021sik, Drukker:2022pxk} (associated to the breaking of the bulk internal symmetries). The AdS background allows to generalize this statement: even for a massive theory in the bulk, the spontaneously broken conserved currents of bulk internal symmetries lead to boundary tilt operators~\cite{Copetti:2023sya}. Via the state/operator correspondence, they can be thought of as the AdS avatars of asymptotic states of Goldstone bosons.\footnote{That genuine Goldstone bosons can exist in two worldsheet dimensions, evading the Coleman-Mermin-Wagner theorem, is a hallmark of the improved infrared behaviour of AdS \cite{Carmi:2018qzm}.}

The open superstring dual to the $1/2$-BPS Wilson-Maldacena line of $\mathcal{N}=4$ super Yang Mills (SYM)~\cite{Maldacena:1998im} provides a fruitful laboratory to realize these ideas. At the leading order in the large $N$ expansion, observables of this line defect can be computed in the holographic dual by considering a bulk fundamental string with a fixed worldsheet, decoupled from the bulk, that ends at the boundary on the defect insertion. The geometry of this classical worldsheet is AdS$_2$, embedded in AdS$_5\times S^5$ and localized at a point on $S^5$.  The Green-Schwarz (GS) superstring $\sigma$-model expanded around this minimal surface is, after static-gauge and $\kappa$-symmetry fixing, an interacting QFT on AdS$_2$ in which the boundary correlators of the $8+8$ worldsheet fluctuations compute correlators of operator insertions on the line. The peculiarity of this theory in AdS$_2$ is that, thanks to the realization as DCFT$_1$ in the boundary $\mathcal{N}=4$ SYM theory, we have access to a complementary point of view on its boundary observables. In particular, on the line there is a protected sector of displacement/tilt operators, that can be seen either as a consequence of the symmetry-breaking due to the line defect in $\mathcal{N}=4$, or as byproducts of SSB in the field theory on AdS$_2$.\footnote{A similar connection between protected displacement operators and worldsheet Goldstone modes has been applied recently to the confining flux tubes in Yang-Mills theory on AdS background~\cite{Gabai:2025jcz,Gabai:2026ads}, see also~\cite{Bianchi:2026sax}.} Moreover, the boundary $1$d correlators enjoy a large supergroup of symmetries~\cite{Liendo:2016ymz, Giombi:2017cqn}, and many of its observables are known exactly from supersymmetric localization~\cite{Correa:2012at,Fiol:2012sg} or integrability. This setup therefore offers a rare situation in which a nontrivial QFT in AdS can be compared, order by order, with first-principles field-theory predictions.

Realizing this comparison beyond the leading semiclassical order is, however, subtle.  The GS action on the $\mathrm{AdS}_5\times S^5$ background is a highly non-linear two-dimensional sigma model whose perturbative quantization requires a regularization prescription compatible with the residual symmetries of the gauge-fixed worldsheet theory.\footnote{Starting at two loops, further structural issues appear. Formal power counting permits higher-derivative local counterterms and associated scheme choices, see discussion in~\cite{Beccaria:2025xry, beccaria2026} and references therein.}
Already in the case of flat induced metric, finding a manifest cancellation of UV divergences is a non-trivial task requiring a particular regularization prescription and scheme~\cite{Roiban:2007jf,Roiban:2007dq}. In a curved-AdS$_2$ setting the situation is even more delicate,  as expected on general grounds~\cite{Giombi:2017hpr,Bertan:2018afl,Liu:2018jhs}.  A cluster of recurring technical questions appear: among others, how to regularize while preserving the explicit and hidden symmetries of the model, and how to consistently define and normalize composite and boundary operators~\cite{Beccaria:2017rbe, Beccaria:2018ocq, Beccaria:2019dws, Beccaria:2021rmj, Beccaria:2022bcr, Beccaria:2019stp, Beccaria:2019mev, Beccaria:2019dju, Beccaria:2020qtk}. 
These difficulties on how to disentangle scheme-dependent artefacts from physical data have so far limited the number of genuine subleading strong-coupling checks performed directly in the worldsheet theory. A sample of previous direct holographic calculations of defect correlators on the Wilson-Maldacena line using the AdS$_2$ string worldsheet, and at leading order in the strong-coupling
expansion, can be found in \cite{Giombi:2017cqn, GiombiKomatsu2018, Bliard:2022zjs, Giombi:2022pas, Giombi:2023mgq}, with analogous studies in other holographic backgrounds discussed in~\cite{Bianchi:2020hsz,Bliard:2024bcz}.

In this paper, our strategy is to focus on the protected subsector associated to SSB mentioned above. This allows us to identify an observable such that: (i) its value is known exactly thanks to localization, and in particular there is a definite prediction to match at subleading order; (ii) we can unambiguously fix its normalization when expressed in terms of the AdS$_2$ degrees of freedom, order by order in the strong coupling expansion. The observable with these properties is the coefficient $C_t(\lambda)$,  function of the 't Hooft coupling $\lambda=N\,g^2_\text{\tiny YM}$,  in the two-point correlation function of the DCFT$_1$ tilt operators
\begin{equation}\label{Ct-intro}
\langle t^p(\tau) t^q(\tau')\rangle = \frac{C_t(\lambda)\delta^{pq}}{|\tau-\tau'|^2}~\,.
\end{equation}
Here the index $p=1,\dots,5$ spans the broken generators $M^{p6}$ of the $SO(6)$ R-symmetry, broken to the $SO(5)$ subgroup by the scalar coupling on the straight Wilson-Maldacena line 
\begin{equation}
    \label{eq:W_intro}
W=\text{tr}Pe^{\,\int d\tau(i\,A_\tau+\Phi^6)}~,
\end{equation}
where $A$ and $\Phi^6$ are the gauge field and one of the scalars of $\mathcal{N}=4$ SYM, respectively. Coordinates on the line are denoted as $\tau,\tau'\in(-\infty,+\infty)$ while $x_\perp^i$ with $i=1,2,3$ are the orthogonal directions. 
The normalization of the tilt operator is fixed by the broken Ward-Takahashi identity of the $SO(6)$ current $\mathcal{J}^p$ in $\mathcal{N}=4$ SYM, namely
\begin{equation}\label{eq:bWTI}
\partial^\mu \mathcal{J}_{\mu}^{p6}(\tau,x_\perp^i)=\delta^{3}(x_\perp^i)\,t^p(\tau)~,
\end{equation}
or -- equivalently -- by the defect operator expansion
\begin{equation}\label{eq:DOEJ1}
\mathcal{J}^p_i(x_\perp, \tau) \underset{x_\perp \to \,0}{\sim} \frac{x_{\perp\,i}}{|x_\perp|^3} \, t^p(\tau) + \dots~,
\end{equation}
making the coefficient $C_t(\lambda)$ in~\eqref{Ct-intro} physical. The identity
\begin{equation}\label{eq:CtBrel}
 C_t(\lambda)=2\,B(\lambda)~,
\end{equation}
relates it to the Bremsstrahlung function $B(\lambda)$ controlling the energy radiated by an accelerated quark, the small-angle cusp anomalous dimension and the displacement two-point function \cite{Correa:2012at}.  
Note that \eqref{eq:bWTI} and \eqref{eq:DOEJ1} imply that the tilt operator has protected dimension~$\Delta = 1$.

In the gravity dual, the SSB of $SO(6)$ down to $SO(5)$ manifests itself in the presence of five~\emph{massless} fields y$^p$ on the string worldsheet. The latter has geometry AdS$_2 \times \text{point}$, embedded in AdS$_5\times $S$^5$. The fields $\mathrm y^p$ are Goldstone bosons that parametrize the fluctuations of the point inside S$^5$. It is then natural to think of $t^p$ as the operator at the boundary of AdS$_2$ that is ``dual'' to the massless field y$^p$. At the level of boundary operator expansion (BOE) the relation is
\begin{equation}\label{eq:bcy_intro}
\mathrm y^p(z,\tau)\underset{z\to 0}{\sim} z \,b_{\mathrm{y} t} \,t^p(\tau) + \dots~.
\end{equation}
Here $z$ and $\tau$ are (Euclidean) AdS Poincaré coordinates (see~\eqref{eq:EAdS2} below)  and $b_{\mathrm{y} t}$ is a BOE coefficient. 
The arbitrary relative normalization $b_{\mathrm{y} t}$ can get corrected beyond the leading linearized approximation. For this reason, without a more principled AdS$_2$ definition of $t^p$, superseding the naive ``dual'' to $\mathrm{y}^p$, it seems difficult to set up a calculation of the subleading strong-coupling correction to $C_t(\lambda)$ purely from the worldsheet theory. This normalization issue is a manifestation of the difficulties in the worldsheet calculations at subleading orders that we mentioned above. 

We resolve this issue by focusing on the spontaneous symmetry breaking of the $SO(6)$ symmetry down to $SO(5)$ directly in the effective AdS$_2$ theory. In this theory one can write down conserved currents $J^p_\alpha(z,\tau)$\footnote{Not to be confused with the current $\mathcal{J}^p_i(x_\perp, \tau)$ in the dual $\mathcal{N}=4$ SYM at the boundary of AdS$_5\times S^5$.} for the broken generators via the Noether procedure, which ensures that they are normalized correctly to satisfy the Ward-Takahashi identity. We then define $t^p$ as the tilt operator for the effective theory in AdS$_2$, namely as the scalar operator appearing in the BOE of the current as follows
\begin{equation}\label{eq:BOEJ1}
J^p_z(z,\tau)\underset{z\to 0}{\sim
} t^p(\tau) + \dots
\end{equation}
where the dots denote terms that are suppressed in the small $z$ expansion.  This BOE expansion is totally analogous to the DOE expansion \eqref{eq:DOEJ1} in $\mathcal{N}=4$ SYM: just like the latter expresses the breaking of the bulk currents $\mathcal{J}^p_i$ of the 4d  CFT due to the line defect,~\eqref{eq:BOEJ1} expresses the breaking of the AdS$_2$ currents $J^p_\alpha$ due to the boundary. 

We show that, rather than from the boundary two-point function of $\mathrm{y}^p$, the coefficient $C_t$ can be extracted more conveniently  from the bulk two-point function of the conserved current, when expressed using the spectral representation \cite{Costa:2014kfa}. The latter, being analogous to a Fourier decomposition, often leads to simplifications in loop calculations, thanks to its simple behavior under convolutions. Using this approach we can show the finiteness of the subleading correction to $C_t$.

Having canceled the divergences, we can test whether the finite result agrees with the relation \eqref{eq:CtBrel}. This requires a careful choice of scheme, to ensure that supersymmetry is preserved. We show that the naive dimensional reduction (DRED) of the worldsheet fails, and must be complemented by a continuation of the number of $S^5$ scalars to $N_\mathrm{y} = 5+2\epsilon$ to recover the correct finite part. This prescription is closely related to the DRED scheme applied to the boundary $\mathcal N=4$ SYM theory~\cite{Erickson:2000af}, in which the continuation of the spacetime dimension $D=4-2\epsilon$ is accompanied by a continuation of the number of scalar fields to $6+2\epsilon$.
With this prescription, the finite result gives
\begin{equation}
C_t(\lambda) = \frac{\sqrt{\lambda}}{2\pi^2} - \frac{3}{4\pi^2} + \mathcal{O}\left(\frac{1}{\sqrt{\lambda}}\right)~,
\end{equation}
in agreement with \eqref{eq:CtBrel} and the known result for $B(\lambda)$ from localization. To our knowledge, this is the first holographic match of a subleading strong-coupling correction to $B(\lambda)$ obtained from a calculation of correlators on the straight line.

{Our result provides evidence that this holographically induced
dimensional-reduction prescription correctly captures the finite
normalization of the protected observable considered here. At present,
however, we do not have an intrinsic derivation of the continuation
$N_{\mathrm{y}}=5+2\epsilon$ from the residual supersymmetry of the
gauge-fixed worldsheet theory, nor do we know whether the same
prescription applies to more general observables or holographic
backgrounds.
It would therefore be useful to test the prescription in higher-point
current correlators and in other realizations of AdS/CFT, in particular
for the $\mathrm{AdS}_4\times\mathbb{CP}^3$ string dual to ABJM theory~\cite{ABJM}.
Another important direction is the extension of the present calculation
to higher loop order. In particular, it would be interesting to determine
whether the divergences found at two loops in the Green--Schwarz string
free energy~\cite{beccaria2026} also arise in current two-point functions, or whether
protected current observables display a better ultraviolet behaviour.}\\

The paper proceeds, in Section~\ref{Section:AdS2}, by recalling the quartic AdS$_2$ Lagrangian of the physical fluctuations transverse to the AdS$_2$ minimal surface dual to the half-BPS Wilson-line; the Lagrangian coincides (up to total derivatives) with the one obtained in static gauge in \cite{beccaria2026} (and earlier in \cite{Giombi:2017cqn,Beccaria:2019dws}), and we keep it in a manifestly covariant form that makes the broken $SO(6)\to SO(5)$ symmetry transparent. Section~\ref{Section:Bremsstrahlung} sets up the tilt observable and the symmetry method: the relation \eqref{Ct-intro}, the harmonic-function extraction of $C_t$ from the current correlator, the broken Noether current, and the tree-level check.  We then evaluate the one-loop diagrams, discuss the coincident-point propagators and the regularization. 
Technical details are collected in the Appendices. In addition, Appendix~\ref{appendix:IIA} presents the Type IIA
superstring extension of the quartic $\mathrm{AdS}_2$ fluctuation
Lagrangian, relevant for holographic calculations of correlators on
the half-BPS Wilson line in ABJM theory.

\newpage
\section{Gauge-fixed fluctuation Lagrangian about the $\mathrm{AdS}_2$ minimal surface}
\label{Section:AdS2}
This section summarizes the worldsheet description of the half-BPS straight Wilson line of $\mathcal{N}=4$ super Yang-Mills in the planar strong-coupling limit. 
A pedagogical derivation of the formulas below is provided in Appendix~\ref{Polyakov}, where conventions are also set.

The bulk description of a 1/2-BPS Wilson line happens via a fundamental open string stretched in AdS,  extended from its conformal boundary into the interior along the radial direction~\cite{Drukker:2000ep}. 
With the common AdS$_5$ and $S^5$  radius set to one, the target space metric is
\begin{align}\label{AdS5xS5metric}
    &\mathrm{d}s^2_{10} = \frac{1}{z^2}\left(\eta_{\mu\nu}\mathrm{d}x^\mu \mathrm{d}x^\nu +\dd z^2\right)  + \mathrm{d}\Omega^2_{S^5}\,. 
\end{align}
The string world surface ending on a straight line along the time direction at the boundary is described by 
\begin{equation}
\label{Embedding}
    x^0 =t=  \sigma^0,\qquad z = \sigma^1,
\end{equation}
in terms of the radius $z$ of AdS$_5$ and the time $t=x^0$, 
with  all remaining coordinates taken  
to vanish.  Below, we also use the symbol $\sigma = (\sigma^0,\sigma^1)$ to denote the vector of coordinates.
The induced metric on the world surface is that of AdS$_2$
\begin{equation}\label{eq:AdSPoinc}
    \mathrm{d}s^2_{\text{ind.}} = \frac{1}{z^2} \left(-\mathrm{d}t^{2} + \mathrm{d}z^2\right)\,.
\end{equation}
The fluctuation action is obtained by expanding the type IIB Green–Schwarz action around the classical solution~\eqref{Embedding}. 
Its covariant derivation,  
based on a Riemann-normal-coordinate (RNC) expansion, is given up to quartic order in the fields in Appendix~\ref{Polyakov}. 
The 2d diffeomorphism invariance can be fixed by imposing a static gauge. In static gauge, the worldsheet coordinates ($\sigma^0$ and $\sigma^1$) are identified with the two target-space coordinates spanning the classical surface ($t$ and $z$), and there are no fluctuations in those directions.
The remaining eight transverse bosonic fluctuations, parametrized by Riemann normal coordinates,  split into three fields $\mathrm{x}^{i}$ ($i=1,2,3$), normal to $\mathrm{AdS}_2$ within $\mathrm{AdS}_5$, and five fields $ \mathrm{y}^{p}$ ($p=5,\cdots,9$), tangent to $S^5$ at the point occupied by the classical string. 
To fix $\kappa$-symmetry we adopt the choice~\cite{beccaria2026} of imposing opposite worldsheet chiralities on the two Type IIB spinors. The effect, at this order in fluctuations, is the vanishing of Yukawa-like couplings $ \mathrm{x}\bar\theta\theta, \mathrm{y}\bar\theta\theta$, which highly simplifies the diagrammatic analysis. 
The expansion of the GS action reads then, up to quartic order in the fields, 
\begin{align}
S &= T\int \dd^2\sigma \sqrt{-g}\,
\bigl(
\mathcal{L}_B^{(2)}
+\mathcal{L}_B^{(4)}
+\mathcal{L}_F^{(2)}
+\mathcal{L}_F^{(4)}
+\mathcal{O}(\theta^4)
+\ldots
\bigr)\,,
\\
\mathcal{L}_B^{(2)}
&= -\frac12\bigl(
g^{\alpha\beta}\partial_\alpha\mathrm{x}^i\partial_\beta \mathrm{x}_i
+g^{\alpha\beta}\partial_\alpha \mathrm{y}^p\partial_\beta \mathrm{y}_p
+2\mathrm{x}^i \mathrm{x}_i
\bigr)\,,
\\
\mathcal{L}_B^{(4)}
&= -\frac12 \mathrm{x}^4
-\frac14 g^{\alpha\beta}\mathrm{x}^i \mathrm{x}_i\,
\partial_\alpha \mathrm{x}^j \partial_\beta \mathrm{x}_j
+\frac16 R_{mpnq} g^{\alpha\beta}
\mathrm{y}^m \mathrm{y}^n
\partial_\alpha \mathrm{y}^p \partial_\beta \mathrm{y}^q
\nonumber\\
&\qquad
+\frac18
\left(
g^{\mu\alpha}g^{\nu\beta}
+g^{\mu\beta}g^{\alpha\nu}
-g^{\mu\nu}g^{\alpha\beta}
\right)
\nonumber\\
&\qquad\quad\times
\Bigl(
\partial_\mu \mathrm{x}^i\partial_\nu \mathrm{x}_i\,
\partial_\alpha\mathrm{x}^j \partial_\beta \mathrm{x}_j
+2\partial_\mu \mathrm{x}^i\partial_\nu \mathrm{x}_i\,
\partial_\alpha \mathrm{y}^p\partial_\beta \mathrm{y}_p
\nonumber\\
&\hspace{5.2cm}
+\partial_\mu \mathrm{y}^p\partial_\nu \mathrm{y}_p\,
\partial_\alpha \mathrm{y}^q\partial_\beta \mathrm{y}_q
\Bigr)\,,
\displaybreak[4]\\
\mathcal{L}_F^{(2)}
&= \frac{i}{2}\bar{\theta}\slashed{\mathcal{D}}\theta\,,
\\
\mathcal{L}_F^{(4)}
&= \frac{i}{8}
\Bigl[
\bigl(
g^{\alpha\beta}g^{\rho\mu}
-g^{\alpha\rho}g^{\beta\mu}
-g^{\alpha\mu}g^{\rho\beta}
\bigr)
\bigl(
\partial_\rho \mathrm{x}^i\partial_\mu\mathrm{x}_i
+\partial_\rho \mathrm{y}^p\partial_\mu\mathrm{y}_p
\bigr)
+2g^{\alpha\beta}\mathrm{x}^i \mathrm{x}_i
\Bigr]
\bar{\theta}\Gamma_\alpha\mathcal{D}_\beta\theta
\nonumber\\
&\qquad
+\frac{i}{8}
\Bigl[
g^{\alpha\beta}
\bigl(
\partial_\alpha\mathrm{x}^i\partial_\beta\mathrm{x}_i
-\partial_\alpha \mathrm{y}^p\partial_\beta\mathrm{y}_p
\bigr)
+2\mathrm{x}^i \mathrm{x}_i
\Bigr]
\bar{\theta}\Gamma_{123}\theta
\nonumber\\
&\qquad
+\frac{i}{8}g^{\alpha\beta}\bar{\theta}
\Bigl(
-\mathrm{x}^i\partial_\alpha \mathrm{x}^j
\Gamma_\beta\Gamma_{ij}
+\frac12\Gamma_\alpha
\mathrm{y}^p\partial_\beta \mathrm{y}^q
R_{pqmn}\Gamma^{mn}
\Bigr)\theta
\nonumber\\
&\qquad
+\frac{i}{8}\epsilon^{\alpha\beta}\bar{\theta}
\left(
\partial_\alpha\mathrm{x}^i\partial_\beta \mathrm{x}^j\Gamma_{ij}
-\del_\alpha\mathrm{y}^p\del_\beta\mathrm{y}^q\Gamma_{pq}
\right)
\Gamma_{123}\theta\,.
\label{eq:AdS2Lagrangian}
\end{align}
Above, $T=1/(2\pi\alpha')\equiv\sqrt{\lambda}/(2\pi)$ is the effective AdS$_5\times S^5$ string tension and $g_{\alpha\beta}$ is the induced AdS$_2$ metric.  
The fluctuation action above defines an interacting QFT on the fixed $\mathrm{AdS}_2$ worldsheet, with loop
expansion parameter $1/T$. Its quadratic bosonic spectrum consists of three scalars $\mathrm{x}^i$ with
$    m_{\mathrm{x}}^2=2$ and
five massless scalars $\mathrm{y}^p$ with $m_{\mathrm{y}}^2=0$. Their indices refer to an orthonormal basis of transverse directions and are therefore contracted with $\delta_{ij}$ and $\delta_{pq}$. 
$\theta$ denotes the ten-dimensional Majorana–Weyl spinor surviving the $\kappa$-symmetry gauge fixing and containing the 16 independent physical real fermionic degrees of freedom. Throughout the paper, we keep the action in this ten-dimensional notation.  In an appropriate gamma-matrix representation, the quadratic fermionic operator is equivalent to that of eight two-dimensional Majorana fermions with squared mass $m^2_F=1$~\cite{Drukker:2000ep}. $\mathcal{D}_\alpha$ is the pullback to the classical worldsheet of the ten-dimensional generalized covariant derivative acting on the GS fermions. It is defined in terms of the pullback of the ordinary spinor covariant derivative $D_\alpha$ and the RR five-form contribution
\begin{equation}
\mathcal{D}_\alpha =  \big(D_\alpha + \frac{1}{2}\Gamma_\alpha\Gamma_{123} \big)\,,\qquad\qquad
D_\alpha  =  \big(\partial_\alpha +\textstyle\frac{1}{4} \omega_{\alpha mn} \Gamma^{mn} \big )\,.
\end{equation}
The fermionic field has been rescaled so that the quadratic kinetic term agrees with the normalization adopted in Ref.~\cite{beccaria2026}.
 We use $\varepsilon^{01}=1$ for the antisymmetric symbol and~$\epsilon^{\mu\nu} = \frac{\eps^{\mu\nu}}{\sqrt{-g}}$ for the corresponding Levi-Civita tensor. 

Together,  bosonic and fermionic degrees of freedom furnish the worldsheet fluctuation spectrum associated with the $\operatorname{OSp}(4^*|4)$ symmetry preserved by the straight half-BPS Wilson line~\cite{Gomis:2006sb,Liendo:2016ymz}. Its bosonic subgroup $SO(2,1)\times SO(3)\times SO(5)$ acts, respectively, as the isometry group of the $\mathrm{AdS}_2$ worldsheet,  and on the three $\mathrm{x}^i$ and five $\mathrm{y}^p$ modes. Consistently with the AdS/CFT dictionary $m^2=\Delta(\Delta-1)$, in the dual DCFT$_1$ the fields $\mathrm{x}^i$ and  $\mathrm{y}^p$ are associated, respectively, to the displacement operator ($\Delta=2$), and to the tilt operator ($\Delta=1$).
In particular,  the fields $\mathrm{y}^p$ describe fluctuations of the point selected on $S^5$ and realize the broken $SO(6)/SO(5)$ symmetry nonlinearly. Their boundary limits define the protected tilt operators associated with the broken $SO(6)$ currents, which are the observables studied in the following section. 
An equivalent fluctuation action may be written using explicit target-space coordinates, as in Refs.~\cite{Giombi:2017cqn} for the bosonic part and~\cite{beccaria2026} for the fermionic one. The two parametrizations are related by local nonlinear field redefinitions and therefore give the same quadratic spectrum, while the explicit interaction vertices, nonlinear symmetry transformations and local Noether currents differ.  All expressions below use the RNC parametrization.

\section{Bremsstrahlung coefficient at strong coupling  from $C_t$}
\label{Section:Bremsstrahlung}

In this section we use the Lagrangian reviewed in the previous section to compute holographically the coefficient $C_t$ of the two-point function \eqref{Ct-intro} of the tilt operator, at first subleading order at strong coupling. To set up the notation we start by the leading order calculation, then we discuss the universal relation between $C_t$ and the bulk two-point function of the conserved current, and finally we perform the calculation using the latter formulation.
We will work here on \AdSd in Euclidean signature
\begin{equation}\label{eq:EAdS2}
    \mathrm{d}s^2_{\text{E}} = \frac{1}{z^2} \left(\mathrm{d}\tau^{2} + \mathrm{d}z^2\right)\,,
\end{equation}
denoting  coordinates jointly by $\sigma = (\sigma^0, \sigma^1)$, with $\sigma^0 = \tau$.

\subsection{$C_t$ at leading order}\label{sec:CtLO}

It is natural to think of $t^p$ as the operator at the boundary of AdS$_2$ that is ``dual'' to the massless field $\mathrm y^p$. At the level of boundary operator expansion (BOE) the relation is
\begin{equation}\label{eq:bcy}
\mathrm{y}^p(z,\tau)\underset{z\to 0}{\sim} z \,b_{\mathrm{y} t} \,t^p(\tau) + \dots~.
\end{equation}
Here $z$ and $\tau$ are (E)AdS Poincaré coordinates  \eqref{eq:EAdS2} and $b_{\mathrm{y} t}$ is a BOE coefficient. Suppose we neglect interactions $\mathcal{L}^{(4)}_B + \mathcal{L}^{(4)}_F$ in \eqref{eq:AdS2Lagrangian} and just consider the linearized theory $\mathcal{L}^{(2)}_B + \mathcal{L}^{(2)}_F$. Then the bulk-to-bulk propagator of $\mathrm y^p$ coming from the quadratic Lagrangian $\mathcal{L}_B^{(2)}$ is 
\begin{equation}\label{eq:yy2pt}
 \langle \mathrm y^p(\sigma)\mathrm  y^q(\sigma')\rangle  = \frac{1}{T} \delta^{pq} \,G_1(\sigma, \sigma')~,
\end{equation}
where
\begin{equation}\label{eq:ScaProp}
G_\Delta(\sigma, \sigma')  \equiv \frac{\Gamma(\Delta)}{2\pi^{\frac{d}{2}} \Gamma(\Delta -\tfrac d2 +1)} \frac{1}{\zeta^{\Delta}} {}_2F_1\left(\Delta,\Delta-\tfrac{d-1}{2},2\Delta-d +1,-\frac{4}{\zeta}\right)~,
\end{equation}
is the standard propagator in AdS$_{d+1}$ for a scalar field with mass-squared $\Delta(\Delta-d)$ and boundary condition $\sim z^{\Delta}$, and
\begin{equation}
\zeta \equiv \frac{(z-z')^2 + (\tau-\tau')^2}{z z'}~,
\label{eq:chordal_distance}
\end{equation}
is the chordal distance between the two points. For $\Delta = d = 1$ it simplifies to 
\begin{equation}\label{eq:yy2ptExpli}
G_1(\sigma, \sigma') = \frac{\log \left(1 +\frac{4}{\zeta}\right)}{4 \pi }~.
\end{equation}
From \eqref{eq:yy2pt}-\eqref{eq:yy2ptExpli} we can compute the two point function of the tilt by sending $z,z' \to 0$, i.e. $\zeta\to\infty$, and stripping off the power $(b_{\mathrm{y} t})^2 z z'$ as prescribed by the boundary condition \eqref{eq:bcy}. This gives
\begin{equation}
\langle t^p(\tau) t^q(\tau')\rangle = \frac{\delta^{pq}}{\pi T \,(b_{\mathrm{y} t})^2\,|\tau-\tau'|^2}(1+\mathcal{O}(T^{-1}))~.
\end{equation}
Upon the identification $T = \frac{\sqrt{\lambda}}{2
\pi}$, the leading coefficient in the large-$\lambda$ expansion of $C_t(\lambda)$ is
\begin{equation}
C_t(\lambda) = \frac{2}{\sqrt{\lambda} \,(b_{\mathrm{y} t})^2 }\left(1+\mathcal{O}\left(\frac{1}{\sqrt{\lambda}
}\right)\right)~.
\end{equation}
Comparing with the answer known from localization
\begin{equation}
C_t(\lambda) = 2 B(\lambda) = \frac{\sqrt{\lambda}}{2 \pi^2}\left(1+\mathcal{O}\left(\frac{1}{\sqrt{\lambda}
}\right)\right)~,
\end{equation}
we obtain that $(b_{\mathrm{y} t})^2 = \frac{1}{T^2} = \frac{4\pi^2}{\lambda}$.

\subsection{Relation between $C_t$ and the current two-point function}\label{sec:C1fromJJ}

To compute $C_t$ at the next subleading order, a natural continuation of the previous section would be to set up the calculation of the boundary two-point function of the field $\mathrm{y}^p$ at one loop. This approach however meets an immediate difficulty: as we saw above there is a BOE coefficient $b_{\mathrm{y}t}$ relating the bulk field to the tilt operator, and this coefficient might receive corrections at subleading order, possibly even UV divergent ones.

As explained in the introduction, an alternative approach is to define the tilt from the BOE of the broken $SO(6)$ current as in AdS$_2$, as follows
\begin{equation}\label{eq:BOEJ}
J^p_z(z,\tau)\underset{z\to 0}{\sim
} t^p(\tau) + \dots
\end{equation}
where the dots denote terms that are suppressed in the small $z$ expansion. 
We perform the detailed Noether calculation of the current operator in the Appendix \ref{app:variation_noether}, and the result is shown in \eqref{eq:currentbos}-\eqref{eq:currentfer} in the next section. As is standard with SSB, the current at leading order is linear in the Goldstone boson fields $\mathrm y^p$. The coefficient is given by
\begin{equation}
J^p_\mu = - T \partial_\mu \mathrm{y}^p + \dots~,
\end{equation}
where the dots denote non-linear terms. Already with the linearized result we can check, using \eqref{eq:bcy} and \eqref{eq:BOEJ}, that indeed at leading order $b_{\mathrm{y} t} = - \frac{1}{T}$, consistently with the result of the previous section.

This definition of $t^p$ in terms of the BOE \eqref{eq:BOEJ} allows us to sidestep the calculation of $b_{\mathrm{y} t}$ at subleading order. This is because the coefficient $C_t(\lambda)$ can actually be fixed in terms of the bulk two point correlator of the current $J_\mu^p$. Let us now open a brief aside and explain how this works.\footnote{See also \cite{DiPietro:2026boh} for an application of this result to 3d gauge theories.} We consider the more general case of a theory in AdS$_{d+1}$ with $d$ arbitrary, with a spontaneously broken bulk conserved current $J_\mu$ satisfying the appropriate generalization of \eqref{eq:BOEJ}, namely
\begin{equation}\label{eq:BOEJd}
J^p_z(z,s)\underset{z\to 0}{\sim
} z^{d-1}\,t^p(s) + \dots~.
\end{equation}
Here $s$ is a $d$ dimensional vector of coordinates for the boundary $\mathbb{R}^d$, and we keep denoting by $\sigma=(z,s)$ the full set of coordinates of a bulk point. Consider the {\it spectral representation} of the two-point function, i.e. its expansion in terms of spin 1 AdS harmonic functions\cite{Costa:2014kfa}\footnote{See also the appendix of \cite{Ankur:2023lum} for the application of AdS harmonic analysis to current two-point functions. More generally the two point function of a vector operator $\langle V_
\mu V_\rho \rangle$ in AdS$_{d+1}$ admits the spectral representation
\begin{equation}
\langle V_\mu(\sigma) V_\rho(\sigma')\rangle = \int_{-
\infty}^{+\infty} d\nu \, B^{\perp}
_V(\nu) (\Omega^{(1)}_\nu)_{\mu\rho}(\sigma, \sigma') + \nabla_\mu \nabla'_\rho\int_{-
\infty}^{+\infty} d\nu \, B^{L}
_V(\nu) \Omega_\nu(\sigma, \sigma')~,
\end{equation}
in terms of two functions $B^{\perp}
_V(\nu)$ and $B^{L}
_V(\nu)$, where $\Omega_\nu$ is the scalar harmonic function. Requiring that the two point function is conserved at separated points gives the constraint $B^{L}
_V(\nu) = c(\nu^2 +\frac{d^2}{4})^{-1}$, with $c$ an arbitrary constant. Further requiring that the conservation holds also at coincident points gives $c=0$.
} 
\begin{equation}\label{eq:specrepJ}
\langle J^p_\mu(\sigma) J^q_\rho(\sigma')\rangle = \delta^{pq}\int_{-
\infty}^{+\infty} d\nu \, B_J(\nu) (\Omega^{(1)}_\nu)_{\mu\rho}(\sigma, \sigma')~.
\end{equation}
We can rewrite the expansion in terms of Proca propagators
\begin{equation}
\Omega^{(1)}_\nu = \frac{i \nu}{2\pi} \left(G^{(1)}_{\frac{d}{2}+i \nu} - G^{(1)}_{\frac{d}{2}-i \nu}\right)~,
\end{equation}
where $G^{(1)}_{\Delta}$ denotes the propagator for a massive free vector dual to a spin 1 operator of scaling dimension $\Delta$. We then separate the integral in two terms, involving $G^{(1)}_{\frac{d}{2}\pm i \nu}$, and close the contour in the complex $\nu$ plane with an arch at infinity in the lower- $(+)$ or upper- $(-)$ half plane. 
Picking the poles at positions $\pm\nu_* = \pm i x$ with $x>0$ of the function $B_J(\nu)$ we obtain discrete contributions to the two point function proportional to $G^{(1)}_{\frac{d}{2}-i \nu_*}$. These are interpreted as the contributions of spin 1 boundary primary operators of dimension $\Delta = \frac{d}{2}-i \nu_*$ appearing in the BOE of $J_\mu$. Moreover, there is a contribution from a simple pole of the Proca propagator\footnote{This equation corrects a sign mistake in equation (54) of \cite{Copetti:2023sya}.}
\begin{equation}\label{eq:Procapole}
\pm \frac{i \nu}{2\pi} \,\left(G^{(1)}_{\frac{d}{2}\mp i \nu}\right)_{\mu\rho} \underset{\pm\nu \to i(\frac{d}{2}-1)}{\sim} \frac{1}{2\pi i} \frac 12 \frac{1}{\pm\nu - i (\frac{d}{2}-1)} \nabla_\mu \nabla'_\rho \,G_d~,
\end{equation}
where, like above, $G_d$ is the propagator of a massless scalar field, and $\nabla$, $\nabla'$ denote derivatives with respect to the points $\sigma$ and $\sigma'$, respectively. These poles give precisely the contribution in the BOE of a scalar boundary primary operator of scaling dimension $\Delta = d$, i.e. the tilt. Therefore they are present only when the symmetry is spontaneously broken. Isolating the contribution from this pole, we can write it as
\begin{equation}\label{eq:BOEtilt2pt}
\langle J_\mu(\sigma) J_\rho(\sigma')\rangle \supseteq - B(i(\tfrac{d}{2}-1)) \nabla_\mu \nabla'_\rho \,G_d(\sigma, \sigma')~.
\end{equation}
We learn that in order to cancel this pole we need to have a zero of the function $B(\nu)$ precisely at $\nu = i(\tfrac{d}{2}-1)$. This is the condition on $B(\nu)$ for the symmetry to be unbroken by the boundary condition. When instead $B(i(\tfrac{d}{2}-1))\neq 0$, by expanding $G_d$ for large $\zeta$ and using the BOE \eqref{eq:BOEJd} we obtain
\begin{equation}
\langle t^p(s) t^q(s')\rangle = - B(i(\tfrac{d}{2}-1)) \frac{\Gamma(d+1)}{\pi^{\frac{d}{2}}\Gamma(\tfrac{d}{2})} \frac{1}{|s-s'|^{2d}}~,
\end{equation}
from which we read off the relation
\begin{equation}\label{eq:generaldCtB}
C_t = -\frac{\Gamma(d+1)}{\pi^{\frac{d}{2}}\Gamma(\tfrac{d}{2})}  B(i(\tfrac{d}{2}-1)) ~.
\end{equation}

For our case of interest $d=1$ this relation becomes
\begin{equation} \label{eq:ourCtB}
C_t = - \frac{1}{\pi} B(-\tfrac{i}{2}) ~.
\end{equation}
Some remarks are in order regarding the specification to $d=1$ of the general result \eqref{eq:generaldCtB}. First, we note that as we continue $d$ from the range $>2$ to $<2$ the pole of the Proca propagator \eqref{eq:Procapole} crosses the integration contour. For $d<2$, if we compute the integral along the real-axis contour as in \eqref{eq:specrepJ}, separating the $G^{(1)}_{\frac{d}{2} + i \nu}$/$G^{(1)}_{\frac{d}{2} - i \nu}$ terms and closing them in the lower/upper-half plane, instead of the pole in \eqref{eq:Procapole} we pick a different pole of the Proca propagator, namely
\begin{equation}\label{eq:ProcapoleAlt}
\pm\frac{i \nu}{2\pi} \,\left(G^{(1)}_{\frac{d}{2}\mp i \nu}\right)_{\mu\rho} \underset{\pm\nu \to -i(\frac{d}{2}-1)}{\sim} \frac{1}{2\pi i} \frac 12 \frac{1}{\pm\nu + i (\frac{d}{2}-1)} \nabla_\mu \nabla'_\rho \,G_0~.
\end{equation}
Here $G_0$ is the finite term in the limit $\Delta\to 0$ of the scalar propagator \eqref{eq:ScaProp} (the divergent part is a constant and drops when acted upon by the derivative operator $\nabla_\mu \nabla'_\rho$). This contribution would correspond to a non-unitary boundary scalar operator of scaling dimension $\Delta = 0$ and therefore it is unphysical. However analyticity in $d$ requires that as $d$ crosses $2$ we deform the contour in \eqref{eq:specrepJ} by adding two infinitesimal circles around the poles in $\pm i (\frac{d}{2}-1)$ with orientation $\mp$ (in the convention that orientation $+$ means counterclockwise). The contribution of the additional circles comes from the single pole of $\Omega^{(1)}_\nu$, namely
\begin{equation}\label{eq:ProcapoleAlt}
\left(\Omega^{(1)}_\nu\right)_{\mu\rho} \underset{\nu \to \pm i(\frac{d}{2}-1)}{\sim} \pm \frac{1}{2\pi i} \frac 12 \frac{1}{\nu \mp i (\frac{d}{2}-1)} \nabla_\mu \nabla'_\rho \,(G_0 - G_d)~.
\end{equation}
Taking into account the orientation, we see that these additional terms precisely cancel the unphysical contribution \eqref{eq:ProcapoleAlt}, giving instead a term of the form \eqref{eq:BOEtilt2pt} like for $d>2$. The fact that a modified contour is needed for the spectral representation of broken currents in $d+1=2$ signals that in this case the square-integrability property needed for the existence of an ordinary (i.e. with real line contour) spectral representation is not satisfied. This is reminiscent of the spectral representation for propagators of fields with ``alternate'' boundary conditions, which also includes similar circles in addition to the real line contour.

\subsection{Broken symmetry  worldsheet currents}

To second order in normal coordinates, the non-linear variations of the worldsheet fields look like \footnote{These results can be understood in terms of coset-space geometry;
see, e.g., Ref.~\cite{Castellani:1999fz} and references therein.}
\begin{equation}
\begin{aligned}
\delta\mathrm{y}^p
&= \varepsilon^p
-\frac{1}{3} R^p{}_{mqn}\mathrm{y}^m\mathrm{y}^n\varepsilon^q
+ o(\mathrm{y}^2)\,,
\\
\delta\theta
&= -\frac{1}{8}\eps^p \mathrm{y}^q R_{pqmn}
\Gamma^{mn}\theta\,,
\\
\delta\bar{\theta}
&= \frac{1}{8}\eps^p \mathrm{y}^q R_{pqmn}
\bar{\theta}\Gamma^{mn}\,.
\end{aligned}
\label{eq:coset_variation}
\end{equation}
where $R^p{}_{mqn}$ is the Riemann tensor of the background spacetime and $\eps^p$ an infinitesimal parameter.
As a check, the Lagrangian (\ref{eq:AdS2Lagrangian}) is shown to be invariant under this transformation in Appendix \ref{app:variation_noether}. The $SO(6)$ broken current reads in Lorentzian signature
\begin{align}
J^\mu_p & = J^\mu_{p,B} +J^\mu_{p,F}~,\\
\frac1T J^\mu_{p,B} &= -\partial^\mu \mathrm{y}_p
+\frac23 R_{pqrs}\del^\mu \mathrm{y}^r \mathrm{y}^q \mathrm{y}^s \nonumber \\
&\qquad
+ g^{\rho\sigma}\bigl(
(\partial^\mu \mathrm{x}^i\partial_\rho \mathrm{x}_i + \partial^\mu \mathrm{y}^q\partial_\rho y_q)\partial_\sigma y_p
-\tfrac12(\partial_\rho \mathrm{y}^q\partial_\sigma \mathrm{y}_q + \partial_\rho \mathrm{x}^i\partial_\sigma \mathrm{x}_i)\partial^\mu \mathrm{y}_p
\bigr),\label{eq:currentbos}\\
\frac1T J^\mu_{p,F}
&=
\frac{i}{4}(g^{\rho\nu}g^{\mu \sigma}- g^{\rho\mu}g^{\nu \sigma}-g^{\rho\sigma}g^{\mu \nu}) \partial_\sigma \mathrm{y}_p\,\bar{\theta}\Gamma_\rho \mathcal{D}_\nu\theta \nonumber \\
&\qquad
-\frac{i}{4}\partial^\mu \mathrm{y}_p\,\bar{\theta}\Gamma_{123}\theta
-\frac{i}{8}\mathrm{y}^q R_{pqmn} \bar{\theta}\Gamma^\mu\Gamma^{mn}\theta - \frac{i}{4}\epsilon^{\mu\nu}\del_\nu \mathrm{y}^q \bar\theta \Gamma_{pq}\Gamma_{123}\theta\,, \label{eq:currentfer}
\end{align}
where we separated into pure bosonic and mixed (bosonic-fermionic) parts for clarity.

    \subsection{Current two-point function and diagrams}
    
In this subsection, we start by re-deriving the leading order result for $C_t$ using the two-point function of the current. We then show the one-loop diagrams that contribute to the next-to-leading result, which are all of tadpole type. Consequently, we will then discuss the computation of tadpole diagrams in Euclidean AdS$_2$.

\paragraph{ Leading order $C_t$ from the current two-point function}
The two-point function of the current at tree level only consists of a single Wick contraction 
\begin{equation}
\label{eq:leading_JJ}
    \nptf{J^p_\mu(\sigma) J^q_\rho(\sigma')}\vert_{\text{tree}} =
    \begin{tikzpicture}
      \begin{feynman}
        \vertex [dot,label=\(-\del_\mu \mathrm{y}^p\)](a) {};
        \vertex [dot,label=\(-\del'_\rho \mathrm{y}^q\)](b) [right=of a]{};
        \diagram* {
          (a) -- [scalar] (b),
        };
      \end{feynman}
    \end{tikzpicture} = T\nptf{\del_\mu \mathrm{y}^p(\sigma
    ) \del_\rho \mathrm{y}^q(\sigma'
    )} =  T\del_\mu \del'_\rho G_d(\sigma, \sigma')\delta^{pq}.
\end{equation}
We are momentarily keeping $d$ generic for future convenience. Because of the normalization of the action (\ref{eq:AdS2Lagrangian}), propagators come with a $1/T$ factor, and insertions of current operators (or of vertices) with a factor of $T$.
Note that, as it is written, this two point function does not satisfy the Ward-Takahashi identity, which prescribes that the divergence of the two-point function should be zero even at coincident points. This is because taking a divergence we get 
\begin{equation}
\label{eq:no_WT}
\del^\mu\del_\mu \del'_\rho G_d(\zeta) = \del'_\rho \Box G_d(\zeta) = - \del'_\rho \delta^{d+1}(\sigma, \sigma')~,
\end{equation}
that is nonvanishing at coincident points. However this can be amended by correcting the two-point function by a term that is itself just a contact term
\begin{equation}
\label{eq:contact_term}
\nptf{J^p_\mu(\sigma) J^q_\rho(\sigma')}_{\text{full}} =  \nptf{J^p_\mu(\sigma) J^q_\rho(\sigma')} - T g_{\mu\rho}(\sigma) \delta^{d+1}(\sigma, \sigma')\delta^{pq}~.
\end{equation}
This is the only choice of scheme for this correlator that is compatible with the symmetry.

In order to use \eqref{eq:ourCtB} to compute $C_t$, we need to compute the spectral representation of this two-point function. To this end, we use the identity \cite{Costa:2014kfa}
\begin{equation}
\label{eq:spectral_identity}
\int_{-\infty}^{+\infty} d \nu \, (\Omega_\nu^{(1)})_{\mu\rho}(\sigma,\sigma'
) = -\del_\mu \del'_\rho G_d(\sigma, \sigma') + g_{\mu\rho}(\sigma) \delta^{d+1}(\sigma, \sigma')~.
\end{equation}
We note that the right-hand side is precisely proportional to the tree-level two-point function of the current. The inclusion of the contact term is of course needed to obtain an integral representation in terms of $\Omega_\nu^{(1)}$, because the latter is transverse even at coincident points. As a result, we obtain that at tree-level the spectral representation is just a constant
\begin{align}
\begin{split}\label{eq:Btree}
\nptf{J^p_\mu(\sigma) J^q_\rho(\sigma')}\vert_{\text{full, tree}} & = \int_{-\infty}^{+\infty} d \nu \, B^{pq}(\nu)_{\text{tree}} \,(\Omega_\nu^{(1)})_{\mu\rho}(\sigma,\sigma'
)~. \\
B^{pq}(\nu)_{\text{tree}}& = -T\, \delta^{pq}~.
\end{split}
\end{align}
This is the final answer for the spectral representation of the two-point function of the linearized current $J_\mu^p = -T \partial_\mu \mathrm{y}^p$ in any $d>2$. The continuation to $d=1$ requires to keep track of the poles of $\Omega_\nu^{(1)}$ at $\nu = \pm i(\frac{d}{2}-1)$, that cross the contour at $d=2$ and move to the opposite half plane for $d=1$. As a result, circles around these points need to be added in $d=1$. This is analogous to the discussion in section \ref{sec:C1fromJJ}. 

Specifying to $d=1$, and plugging \eqref{eq:Btree} in \eqref{eq:ourCtB}, we get
\begin{equation}
(C_t)_{\text{tree}} = \frac{T}{\pi} = \frac{\sqrt{\lambda}}{2\pi^2}~,
\end{equation}
in agreement with the direct calculation in section \ref{sec:CtLO} and with the localization result for $B(\lambda)$.

    \paragraph{One loop diagrams}
At one loop, two types of diagrams contribute to the current two-point function: current and tadpole diagrams (see Fig. \ref{fig:example_diagrams}). The current diagrams are the contributions of higher terms in the expansion (\ref{eq:currentbos}) and (\ref{eq:currentfer}) of the current. The tadpoles, on the other hand, are the usual perturbative expansion diagrams arising from the interactions in the quartic terms $\mathcal{L}^{(4)}_B$ and $\mathcal{L}^{(4)}_F$. To evaluate the diagrams, we perform a continuation to Euclidean signature of both the Lagrangian and the currents.
\begin{center}
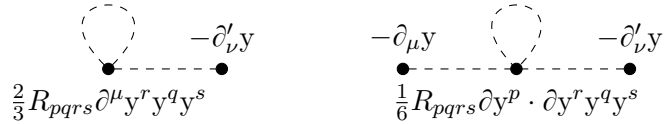

    \begin{tikzpicture}
      \begin{feynman}
        \vertex [dot,label=below:\(\frac23 R_{pqrs}\del^\mu \mathrm{y}^r \mathrm{y}^q \mathrm{y}^s\)](a) {};
        \vertex [dot,label=above:\(-\del'_\nu \mathrm{y}\)](b) [right=of a]{};
        \diagram* {
          (a) -- [scalar] (b),
          a --[scalar, out=135,in=45,loop,min distance=1.5cm] a,
        };
      \end{feynman}
    \end{tikzpicture}
    \hspace{1cm}
    \begin{tikzpicture}
      \begin{feynman}
        \vertex [dot,label=below:\(\frac16 R_{pqrs}\del \mathrm{y}^p \cdot \del \mathrm{y}^r \mathrm{y}^q \mathrm{y}^s\)](a) {};
        \vertex [dot,label=above:\(-\del'_\nu \mathrm{y}\)](b) [right=of a]{};
        \vertex [dot,label=above:\(-\del_\mu \mathrm{y}\)](c) [left=of a]{};
        \diagram*{
          (c) -- [scalar] (b),
          a --[scalar, out=135,in=45,loop,min distance=1.6cm] a,
        };
      \end{feynman}
    \end{tikzpicture}
    \captionof{figure}{Example of current (left) and tadpole (right) diagrams.} 
    \label{fig:example_diagrams}
\end{center}

We note the following simplifications in the diagrams with x or $\theta$ loops: any non-vanishing current diagram is equal to minus its tadpole counterpart, namely
\begin{equation}
    \begin{tikzpicture}[baseline=(current bounding box.center)]
      \begin{feynman}
        \vertex [dot,label=below:$2\del^\rho \mathrm y \mathcal{V}^{(\mathrm{x},\theta)}_{\mu\rho}$](a) {};
        \vertex [dot,label=below:\(-\del'_\nu \mathrm{y}\)](b) [right=of a]{};
        \diagram* {
          (a) -- [scalar] (b),
          a --[out=135,in=45,loop,min distance=1.5cm] a,
        };
      \end{feynman}
    \end{tikzpicture}
    = (-1)\times
    \begin{tikzpicture}[baseline=(current bounding box.center)]
      \begin{feynman}
        \vertex [dot,label=below:$\del^\rho \mathrm y \del^\sigma \mathrm y \mathcal{V}^{(\mathrm{x},\theta)}_{\rho\sigma}$](a) {};
        \vertex [dot,label=above:\(-\del'_\nu \mathrm{y}\)](b) [right=of a]{};
        \vertex [dot,label=above:\(-\del_\mu \mathrm{y}\)](c) [left=of a]{};
        \diagram* {
          (c) -- [scalar] (b),
          a --[out=135,in=45,loop,min distance=1.6cm] a,
        };
      \end{feynman}
    \end{tikzpicture},
    \label{eq:V_diag}
\end{equation}
with $\mathcal{V}^{(\mathrm{x},\theta)}_{\mu\nu}$ a certain interaction term in the Lagrangian. To explain this relation, note that the same structure $\mathcal{V}^{(\mathrm{x},\theta)}_{\mu\nu}$ of the vertex appears also in the current operator. Moreover, the tadpole diagram has two $\mathrm{y}$ propagators while the current diagram has only one. However, the Wick contraction of the derivatives of the $\mathrm{y}$ fields satisfies the convolution identity
\begin{align}
\begin{split}
        \int d^2 \sigma' \sqrt{g} \, \del_\alpha \del'_\rho G_d(\sigma,\sigma'){\del'}^\rho \del''_\beta  G_d(\sigma', \sigma'')    =  \partial_\alpha \partial''_\beta G_d\left(\sigma,\sigma''\right)\,,
\end{split}
    \label{eq:convolution}
\end{align}
which allows to rewrite the integral of two propagators as a single propagator, thus explaining the observed simplification. The convolution identity \eqref{eq:convolution} can be derived using the relation \eqref{eq:spectral_identity} and the convolution identity for the harmonic function
\begin{align}
\begin{split}
        \int d^2 \sigma' \sqrt{g} \, (\Omega_\nu^{(1)})_{\alpha\rho}(\sigma,\sigma')(\Omega_\nu^{(1)})^\rho_{~\beta}(\sigma', \sigma'')
    =  (\Omega_\nu^{(1)})_{\alpha\beta}\left(\sigma,\sigma''\right)\,,
\end{split}
    \label{eq:convolution2}
\end{align}
or directly by integration by parts and the use of the equation of motion \eqref{eq:no_WT}. Note that, for the fermionic loops, we used that only the non-zero contributions come from terms involving the bilinears $\bar \theta \Gamma_\alpha \mathcal{D}_\beta\theta$ and $\bar \theta \Gamma^{123}\theta$, in which $\mathrm{y}$ only enters with a derivative. We will come back to this simplification in Subsection \ref{subsec:results}, where the explicit computation of the relevant diagrams is provided.

    \subsection{Coincident-points limit of propagators}
    \label{subsec_prop}
In this subsection, we provide more details on the computations of the diagrams. We give the results in general dimension, as we will later regulate divergences by (a modified) dimensional regularization. We also give the regularized expressions for the coincident-points limit propagators, as they are needed to compute the tadpoles (see Figure \ref{fig:example_diagrams}).

    \paragraph{Scalar propagators} 
We introduced the propagator of a scalar field in E$\AdS$ in (\ref{eq:ScaProp}), in terms of the chordal distance \eqref{eq:chordal_distance}. By making use of standard properties of the hypergeometric functions,\footnote{See the Digital Library of Mathematical Functions (DLMF) 15.8.1.} it can be rewritten as
\begin{equation}
\label{s_prop}
    G_\Delta(\sigma, \sigma') = \frac{\Gamma(\Delta)}{2\pi^{d/2}\Gamma\!\left(\Delta-\frac d2+1\right)}
    \,(\zeta+4)^{-\Delta}
    \,{}_2F_1\!\left(\Delta,\Delta-\frac{d-1}{2};\,2\Delta-d+1;\,\frac{4}{\zeta+4}\right)\, .
\end{equation}
This expression has the advantage of being manifestly regular at $\zeta=0$ whenever the hypergeometric function is finite for unit argument, namely in the range $\mathrm{Re}(d)<1$. Analytically continuing from this range therefore easily gives us the coincident point limit
\begin{align}
\begin{split}
  G_{\Delta}\big\vert_0 & =\frac{4^{-\Delta}\Gamma(\Delta)}{2\pi^{d/2}\Gamma\!\left(\Delta-\frac d2+1\right)}\,{}_2F_1\!\left(\Delta,\Delta-\frac{d-1}{2};\,2\Delta-d+1;\,1\right) \\
& =\frac{\Gamma(\Delta)\Gamma\!\left(\frac{1-d}{2}\right)}{(4\pi)^{(d+1)/2}\Gamma(\Delta-d+1)}\, .
\end{split}
\end{align}
Here and in the following, we denote with $\big\vert_0$ the evaluation at coincident points $(\sigma, \sigma)$. Note that, since the regulator preserves the spacetime symmetry, the first derivative of the propagator $\partial_\mu G_{\Delta}\big\vert_0$ vanishes at coincident points. On the other hand, the second derivative at coincident points can be proportional to the metric tensor at that point. To compute the proportionality constant, we can use the equations of motion, i.e. the fact that the propagator is a Green function for the Klein-Gordon equation, namely
\begin{equation}
    \Box G_\Delta(\sigma,\sigma') =- \delta^{d+1}(\sigma,\sigma') + m^2 G_\Delta(\sigma,\sigma')\,.
\end{equation}
As a result, we obtain
\begin{equation}
    \partial_\rho\partial'_\sigma G_{\Delta}\big\vert_0
    =
    -\frac{g_{\rho\sigma}}{d+1} m^2 G_{\Delta}\big\vert_0 \,.
    \label{eq:EOMidentity_scalar}
\end{equation}
We used that in dimensional regularization the Dirac delta at coincident points vanishes $\delta^{d+1}(\sigma,\sigma)=0$.

    \paragraph{Fermion propagators} 
Expressions for the fermion propagator in E$\AdS$ are given in many different forms in the literature, but can of course be shown to be all equivalent \cite{Kawano_2000,Muck_2000,Basu_2006,Giombi:2021cnr,beccaria2026}. In our case, since the mass term in the Lagrangian (\ref{eq:AdS2Lagrangian}) is not diagonal, one needs to perform a chiral rotation of the standard expression. This has been performed in \cite{beccaria2026} and the result is\footnote{Here, we use the standard notation $\slashed{\sigma} = \sigma^a \Gamma_a$ with $a=0,... d$ a flat index  \cite{Giombi:2021cnr,beccaria2026}.
}
\begin{equation}
      S(\sigma,\sigma') =
    \frac{-\Gamma\left(\Delta+\frac12\right)}
    {2\pi^{d/2}\Gamma\left(\Delta-\frac{d-1}{2}\right)}
    \frac{(\zeta+4)^{-\Delta-\frac12}}{\sqrt{zz'}}
    \left(
    \Gamma_{123}\bigl(\slashed{\sigma}\Gamma_4+\Gamma_4\slashed{\sigma}'\bigr)F_1(\zeta) - (\slashed{\sigma}-\slashed{\sigma}') F_2(\zeta)\right)\,,
    \label{eq:fermionic_propagator}
\end{equation}
with 
\begin{align}
\begin{split}
\label{F12}
    F_1 &= \,_2F_1\!\left(\Delta+\frac12,\Delta-\frac d2;\,2\Delta-d+1;\,\frac{4}{\zeta+4}\right), \\ F_2 &= \,_2F_1\!\left(\Delta+\frac12,\Delta-\frac d2+1;\,2\Delta-d+1;\,\frac{4}{\zeta+4}\right)\,.
    \end{split}
\end{align} 
In the coincident point limit, the propagator is regular in the range $\mathrm{Re}(d)<1$. Analytically continuing from there, we get 
\begin{equation}
\label{S0}
  S\big\vert_0=\frac{-4^{-\Delta}\Gamma\!\left(\Delta+\frac12\right)}
    {4\pi^{d/2}\Gamma\left(\Delta-\frac{d-1}{2}\right)}
    \,\frac{\Gamma_{123}}{z}\,
    \{\slashed{\sigma},\Gamma_4\}
    \,{}_2F_1\!\left(\Delta+\frac12,\Delta-\frac d2;\,2\Delta-d+1;\,1\right) = S_0 \Gamma_{123}\,,
\end{equation}
with 
\begin{equation}
\label{S00}
    S_0 =-\frac{\Gamma\!\left(\Delta+\frac12\right)\Gamma\!\left(\frac{1-d}{2}\right)}
{(4\pi)^{(d+1)/2}\Gamma\!\left(\Delta-d+\frac12\right)}\,.
\end{equation} 
Unlike the scalar case, now the first derivative at coincident points is not set to zero by the spacetime symmetry because it can be proportional to the gamma matrix $\Gamma_\mu$. 
An explicit computation yields
\begin{equation}
  D_\mu S\big\vert_{0} = \frac{m}{d+1}S_0\,\Gamma_\mu\,.
  \label{eq:EOMidentity_fermion} 
\end{equation}
The proportionality constant is fixed using the equation of the propagator at coincident points $\slashed{D}S = -m \Gamma_{123}S$ (where again we used that in dimreg $\delta^{d+1}(\sigma,\sigma)=0$).  All the results of this section are in agreement with \cite{beccaria2026}.

    \subsection{Results and regularization}
    \label{subsec:results}
\paragraph{Dimensional reduction}\, 
Let us start by detailing our regularization procedure. We use a SUSY-preserving variation of dimensional regularization called dimensional reduction (DRED) \cite{SIEGEL1979193}.
In DRED, vector and spinor indices of both the fields and composite operators, are kept strictly in $d+1=2$ dimensions, while all operators are assumed to depend only on $d+1=2-2\eps$ spacetime coordinates. Derivatives have a special status, being both vectors and closely linked to spacetime coordinates. They should be regarded as strictly $d+1=2$-dimensional objects, but are in practice non-vanishing only along the $2-2\eps$ spacetime dimensions. The implications of this fact will be discussed below. 
 
 The scalar quantities $G_{\Delta}$ and $S_0$ read in $d+1=2-2\eps$
\begin{align}
  G_{\Delta}\big\vert_0 &= \begin{cases}
  \frac{1}{4\pi\eps'} - \frac{1}{2\pi} & \text{for}~\Delta_{\mathrm{x}}=\frac{d}{2}+\sqrt{\frac{d^2}{4}+2}\\
  \frac{1}{4\pi\eps'} & \text{for}~\Delta_{\mathrm{y}}=d
    \end{cases}, & S_0 &= -\frac{1}{4\pi\eps'} + \frac{1}{4\pi} \quad \text{for}~\Delta_\theta=\frac d2+1\,,
  \label{eq:propagators_dimreg}
 \end{align}
where $\frac{1}{\eps'}=\frac{1}{\eps}+\gamma+\mathrm{log}(4\pi)$ absorbs the scheme dependent finite factors. Notice that we keep the $d+1=2-2\varepsilon$-dependence in the scaling dimensions $\Delta$. This is because the relation between $\Delta$, $d$ and the masses of the fields comes from the equations of motion. The parameter $d$ appears when applying derivatives, and is therefore tied to the spacetime coordinates. Moreover, we assume that the masses of the fields are $\varepsilon$-independent when we perform the continuation. Note that with this prescription the dimension of x deviates from $d+1$, the expected value for a displacement operator, at order $\varepsilon$. Changing the value of the mass-squared and of the scaling dimension of x at order $\varepsilon$ does not affect the final result.

The equation of motion (EOM) identities \eqref{eq:EOMidentity_scalar} and \eqref{eq:EOMidentity_fermion} from the previous section contain the parameter $d$, and one needs to check whether in DRED this parameter has to be taken $\varepsilon$-dependent or not. For the bosonic identity \eqref{eq:EOMidentity_scalar}, the metric in the right hand side appears from the identity $\del_\mu \del_\nu \zeta\vert_0 =2g_{\mu\nu}$. Since derivatives are only nonvanishing along $2-2\eps$ dimensions, that metric should be replaced by its $2-2\eps$ dimensional version $\hat g_{\mu\nu}$. This hat-metric has the following properties 
\begin{align}
    \hat g_{\mu\nu}g^{\mu\nu}&=2-2\epsilon\,, & \hat g_{\mu\rho}g^{\rho\nu}&=\hat \delta^\nu_\mu\,,
\end{align}
namely it is a projector in the $2-2\eps$ dimensional space. As a consequence, the factor in the denominator of \eqref{eq:EOMidentity_scalar} is $d+1=2-2\varepsilon$ in DRED. For the fermionic identity \eqref{eq:EOMidentity_fermion}, it is important to track whether the analogous factor $\frac{1}{d+1}$ arises from the gamma matrices, which is strictly 2-dimensional in DRED, or from the spacetime coordinates. One can check that this factor appears as a consequence of recursion relation of the $\Gamma$ functions when relating the function $F_2$ to $S_0$ at coincident points. The parameters of $F_2$ hypergeometric function contain the $\varepsilon$ dependence of $d$, because they are fixed by the equation of motion, which relates them to the spacetime dimensions. Therefore also that factor in DRED contains $d+1=2-2\varepsilon$.\footnote{Note that using $d+1=2$ or $=2-2\epsilon$ for the factor in the fermionic EOM relation actually does not affect the final result of our calculation.\label{foot:Dstrict}}

Notice that the worldsheet gamma matrices $\Gamma_\mu$ satisfy a Clifford algebra that closes with the 2-dimensional $g_{\mu\nu}$ metric. Similarly, the worldsheet Levi-Civita symbol $\varepsilon^{\alpha\beta}$ of the Wess-Zumino terms in the Lagrangian is kept strictly 2-dimensional. In any case, all contributions coming from $\varepsilon^{\alpha\beta}$-terms vanish before any dimensional continuation is required thanks to $\Gamma$'s trace properties.

\paragraph{Diagram list}    We now proceed to compute the diagrams. To avoid weighing down
the presentation, here we provide the main results while a detailed example can be found in Appendix \ref{app:example}. 
    
By using $\partial_\mu G_{\Delta}\big\vert_0 =0$ and the EOM identity (\ref{eq:EOMidentity_scalar}) with $m^2=0$, we observe that any current and tadpole diagrams involving derivatives of the y propagator evaluated at coincident points vanish. This leaves us with only 2 $\mathrm{y}$-loop diagrams in which two fields y without derivatives can close the tadpole, namely
\begin{align}
\begin{split}
        \begin{tikzpicture}[baseline=(current bounding box.center)]
    \begin{feynman}
        \vertex [dot,label=below:\(\frac23 R_{ptrs}\del_\mu \mathrm{y}^r \mathrm{y}^t \mathrm{y}^s\)](a) {};
        \vertex [dot,label=above:\(-\del'_\nu \mathrm{y}_q\)](b) [right=of a]{};
        \diagram* {
          (a) -- [scalar] (b),
          a --[scalar, out=135,in=45,loop,min distance=1.5cm] a,
        };
    \end{feynman}
    \end{tikzpicture} &= -\partial_\mu\partial_\nu^\prime G_d(\sigma,\sigma^\prime)\frac23 R_{pq} G_d\big\vert_0 \,,\\
    \begin{tikzpicture}[baseline=(current bounding box.center)]
    \label{diag_2}
    \begin{feynman}
        \vertex [dot,label=below:\(\frac16 R_{turs}\del \mathrm{y}^t \cdot \del \mathrm{y}^r \mathrm{y}^u \mathrm{y}^s\)](a) {};
        \vertex [dot,label=above:\(-\del'_\nu \mathrm{y}_q\)](b) [right=of a]{};
        \vertex [dot,label=above:\(-\del_\mu \mathrm{y}_p\)](c) [left=of a]{};
        \diagram*{
          (c) -- [scalar] (b),
          a --[scalar, out=135,in=45,loop,min distance=1.6cm] a,
        };
    \end{feynman}
    \end{tikzpicture} &= \partial_\mu\partial_\nu^\prime G_d(\sigma,\sigma^\prime)\frac13 R_{pq} G_d\big\vert_0 \,,
\end{split}
\end{align}
where $R_{pq}$ is the Ricci tensor of the five-sphere, and can be expressed as $R_{pq}=(N_{\mathrm{y}}-1)\delta_{pq}$. $N_{\mathrm{y}}$ (and later $N_{\mathrm{x}}$ and $N_{\theta}$) stands for the number of y (resp. x and $\theta$) degree of freedom. We keep it explicit for future convenience.
When evaluating the tadpole in \eqref{diag_2} we used \eqref{eq:convolution}. For the x-loop diagrams, using the EOM identities \eqref{eq:EOMidentity_scalar}, we obtain the following results
  \begin{align}
     \label{eq:x_diagrams}
\begin{split}
       \begin{tikzpicture}[baseline=(current bounding box.center)]
      \begin{feynman}
        \vertex [dot,label=below:\(\del_\mu \mathrm{x} \cdot \del \mathrm{x} \del \mathrm{y}_p \)](a) {};
        \vertex [dot,label=above:\(-\del'_\nu \mathrm{y}_q\)](b) [right=of a]{};
        \diagram* {
          (a) -- [scalar] (b),
          a --[ghost, out=135,in=45,loop,min distance=1.5cm] a,
        };
      \end{feynman}
    \end{tikzpicture} &= \delta_{pq} \partial_\mu\partial_\nu^\prime G_d(\sigma,\sigma^\prime)\frac{N_{\mathrm{x}}}{(d+1)}m_\mathrm{x}^2 G_{\Delta_{\mathrm{x}}}\big\vert_0 =-
    \begin{tikzpicture}[baseline=(current bounding box.center)]
      \begin{feynman}
        \vertex [dot,label=below:\( \frac12 (\del \mathrm{x} \del \mathrm{y})^2 \)](a) {};
        \vertex [dot,label=above:\(-\del'_\nu \mathrm{y}_q\)](b) [right=of a]{};
        \vertex [dot,label=above:\(-\del_\mu \mathrm{y}_p\)](c) [left=of a]{};
        \diagram* {
          (c) -- [scalar] (b),
          a --[ghost, out=135,in=45,loop,min distance=1.6cm] a,
        };
      \end{feynman}
    \end{tikzpicture} \,,
\\
    \begin{tikzpicture}[baseline=(current bounding box.center)]
      \begin{feynman}
        \vertex [dot,label=below:\(-\frac12 \del_\mu \mathrm{y}_p(\del \mathrm{x})^2\)](a) {};
        \vertex [dot,label=above:\(-\del'_\nu \mathrm{y}_q\)](b) [right=of a]{};
        \diagram* {
          (a) -- [scalar] (b),
          a --[ghost, out=135,in=45,loop,min distance=1.5cm] a,
        };
      \end{feynman}
    \end{tikzpicture} &= -\delta_{pq}\partial_\mu\partial_\nu^\prime G_d(\sigma,\sigma^\prime)\frac{N_{\mathrm{x}}}{2} m_\mathrm{x}^2 G_{\Delta_{\mathrm{x}}}\big\vert_0 \, =-
    \begin{tikzpicture}[baseline=(current bounding box.center)]
      \begin{feynman}
        \vertex [dot,label=below:\( -\frac14 (\del \mathrm{x})^2 (\del \mathrm{y})^2 \)](a) {};
        \vertex [dot,label=above:\(-\del'_\nu \mathrm{y}_q\)](b) [right=of a]{};
        \vertex [dot,label=above:\(-\del_\mu \mathrm{y}_p\)](c) [left=of a]{};
        \diagram* {
          (c) -- [scalar] (b),
          a --[ghost, out=135,in=45,loop,min distance=1.6cm] a,
        };
      \end{feynman}
    \end{tikzpicture}\,.
\end{split}
  \end{align}
In the fermionic sector, several contributions vanish for two reasons. Diagrams containing $\theta$-bilinears with “$\Gamma_{pq}$” factors (see e.g. the last two terms in \eqref{eq:currentfer} or the last two lines of $\mathcal{L}^{(4)}_F$ in \eqref{eq:AdS2Lagrangian}) cancel out when taking the fermionic trace. This is also in agreement with the fact that a non-zero contribution from those terms would be incompatible with the $p,q$ symmetry of $\langle J^p_\mu(\sigma) J^q_\rho(\sigma')\rangle$.  
Additionally, diagrams containing “$\mathcal{D}\theta$” terms (arising e.g. from first lines in $\mathcal{L}^{(4)}_F$ in \eqref{eq:AdS2Lagrangian} and $J^\mu_{p,F}$ in \eqref{eq:currentfer}) vanish thanks to the EOM identity in (\ref{eq:EOMidentity_fermion}), indeed
\begin{equation}
    \langle \bar{\theta} \Gamma_\mu \mathcal{D}_\nu \theta \rangle \big\vert_0
    =-\operatorname{tr}\!\left(
    \frac{m_F}{d+1}\Gamma_\mu\Gamma_\nu + \frac12\Gamma_\mu\Gamma_\nu\Gamma_{123}^2
    \right)S_0
    =-\frac{\eps}{2}g_{\mu\nu}N_\theta S_0\,,
\end{equation}
where we used $m_F=1$. This quantity is then contracted with the traceless tensor $g^{\mu\alpha}g^{\nu\beta}+g^{\mu\beta}g^{\alpha\nu}-g^{\mu\nu}g^{\alpha\beta}$ yielding an order $\eps^2$ contribution.\footnote{To complement on footnote \ref{foot:Dstrict}, if we take $d+1=2$ in EOM identity (\ref{eq:EOMidentity_fermion}), we obtain the same result since
\begin{equation}
    \langle \bar{\theta} \Gamma_\mu \mathcal{D}_\nu \theta \rangle \big\vert_0
    =
    -\operatorname{tr}\!\left(
    \frac{m_F}{2}\Gamma_\mu\Gamma_\nu + \frac12\Gamma_\mu\Gamma_\nu\Gamma_{123}^2
    \right)S_0
    =
    -\operatorname{tr}\!\left(
    \frac12 \Gamma_\mu\Gamma_\nu - \frac12 \Gamma_\mu\Gamma_\nu
    \right)S_0=0\,.
\end{equation}} 
Hence, only two diagrams remain\footnote{Notice that the traces run only on the physical $N_\theta=16$ fermionic degrees of freedom.}
    \begin{equation}
      \begin{tikzpicture}[baseline=(current bounding box.center)]
      \begin{feynman}
        \vertex [dot,label=below:\( \frac{1}{4} \del_\mu \mathrm{y}_p \bar{\theta}\Gamma_{123}\theta \)](a) {};
        \vertex [dot,label=above:\(-\del'_\nu \mathrm{y}_q\)](b) [right=of a]{};
        \diagram* {
          (a) -- [scalar] (b),
          a --[out=135,in=45,loop,min distance=1.5cm] a,
        };
      \end{feynman}
    \end{tikzpicture} = -\delta_{pq}\partial_\mu\partial_\nu^\prime G_d(\sigma,\sigma^\prime)\frac{1}{4}N_\theta S_0\, =-
    \begin{tikzpicture}[baseline=(current bounding box.center)]
      \begin{feynman}
        \vertex [dot,label=below:\( \frac{1}{8} (\del \mathrm{y})^2 \bar{\theta}\Gamma_{123}\theta \)](a) {};
        \vertex [dot,label=above:\(-\del'_\nu \mathrm{y}_q\)](b) [right=of a]{};
        \vertex [dot,label=above:\(-\del_\mu \mathrm{y_p}\)](c) [left=of a]{};
        \diagram* {
          (c) -- [scalar] (b),
          a --[out=135,in=45,loop,min distance=1.6cm] a,
        };
      \end{feynman}
    \end{tikzpicture}\,.
    \label{ferm_diag}
  \end{equation}

Finally, by looking at \eqref{eq:x_diagrams} and \eqref{ferm_diag}, we can observe that in the $\mathrm{x}$ and $\theta$ sectors, the property in \eqref{eq:V_diag} holds.

\paragraph{Sum and finite part}

To perform the diagram sum, the contributions of current diagrams have to be doubled, since the current loop can be attached to either end of the y propagator. The results, given separately for the loops of y, x and $\theta$, are
\begin{align}
\begin{split}\label{eq:resultbyfield}
      \mathrm y\ &: -\partial_\mu\partial_\nu^\prime G_d(\sigma,\sigma^\prime)\delta_{pq}(N_{\mathrm{y}}-1)G_d\big\vert_0 =-\partial_\mu\partial_\nu^\prime G_d(\sigma,\sigma^\prime)\delta_{pq}(N_{\mathrm{y}}-1)\frac{1}{4\pi \varepsilon'}\,,\\
       \mathrm x\ &: -\partial_\mu\partial_\nu^\prime G_d(\sigma,\sigma^\prime)\delta_{pq}\left(\frac{1}{2}-\frac{1}{d+1}\right)N_{\mathrm{x}} m_\mathrm{x}^2 G_{\Delta_{\mathrm{x}}}\big\vert_0 = -\partial_\mu\partial_\nu^\prime G_d(\sigma,\sigma^\prime)\delta_{pq}N_{\mathrm{x}} \left(-\frac{1}{4\pi}\right)\,, \\
    \theta\ &: -\partial_\mu\partial_\nu^\prime G_d(\sigma,\sigma^\prime)\delta_{pq}\frac{N_\theta }{4} S_0 = -\partial_\mu\partial_\nu^\prime G_d(\sigma,\sigma^\prime)\delta_{pq}\frac{N_\theta}{4}\left(-\frac{1}{4\pi \eps'}+\frac{1}{4\pi}\right)\,.  
\end{split}
\end{align}
Plugging the number of fields  $N_\mathrm{y} = 5$ and $N_\mathrm{\theta} = 16$ ($N_\mathrm{x} = 3$ does not play a role in this) we see that the UV divergence cancels in the sum of the three contributions. The cancellation is expected to be universal, i.e. independent of the particular choice of DRED regularization that we used for the explicit calculation. This demonstrates that the approach based on the bulk current allows to reliably compute $C_\mathrm{t}$ in perturbation theory in AdS$_2$. In the next section we will contrast it with a more naive approach based on the computation of the two-point of function of the field $\mathrm{y}$, and show that in that case the UV divergence does not cancel.

Having ensured the cancellation of divergences, we can compute the finite value of $C_\mathrm{t}$. In order to obtain the value that matches the relation to the Bremsstrahlung function \eqref{eq:CtBrel} one needs to work in an appropriate scheme. We observe that using purely DRED for the two-dimensional field theory does not give the expected result. However, the correct answer is obtained if we supplement DRED with a continuation of the number of massless scalar fields to $N_{\mathrm{y}}=5+2\eps$. This is natural from the point of view of the holographic interpretation of the setup, because it amounts to keeping fixed $d+1+N_\mathrm{y}+N_{\mathrm{x}} = 10$. With this modification, plugging in \eqref{eq:resultbyfield}, we find that the finite terms for the contributions of each field are
  \begin{align*}
      \mathrm y\ &: -\partial_\mu\partial_\nu^\prime G_d(\sigma,\sigma^\prime)\delta_{pq}\left( \frac{1}{\pi\eps'}+\frac{1}{2\pi}\right)\,,\notag\\
       \mathrm x\ &:  -\partial_\mu\partial_\nu^\prime G_d(\sigma,\sigma^\prime)\delta_{pq}\left(-\frac{3}{4\pi}\right)\,, \\
    \theta\ &:  -\partial_\mu\partial_\nu^\prime G_d(\sigma,\sigma^\prime)\delta_{pq}\left(-\frac{1}{\pi \eps'}+\frac{1}{\pi}\right)\,,  
  \end{align*}
from which
\begin{equation}
\label{eq:JJoneloop}
    \langle J^p_\mu(\sigma) J^q_\rho(\sigma')\rangle\vert_{\text{one-loop}}= \langle J^p_\mu(\sigma) J^q_\rho(\sigma')\rangle\vert_{\text{tree}}-\frac{3}{4\pi}\del_\mu \del'_\rho G_d(\sigma, \sigma')\delta^{pq}\,.
\end{equation}
Note that the one-loop correction, up to the coefficient, takes the same form as the tree-level contribution in \eqref{eq:leading_JJ}. 

Similarly to what we discussed in \eqref{eq:no_WT}, the naive sum of the tree-level and one-loop contributions to the current two-point functions does not satisfy the Ward-Takahashi identity, and a contact term should be introduced to get around this issue. The quantum-corrected version of \eqref{eq:contact_term} reads
\begin{equation}
\nptf{J^p_\mu(\sigma) J^q_\rho(\sigma')}\vert_{\text{full},\text{one-loop}} =  \nptf{J^p_\mu(\sigma) J^q_\rho(\sigma')}\vert_{\text{one-loop}} - \left(T-\frac{3}{4\pi}\right) g_{\mu\rho}(\sigma) \delta^{d+1}(\sigma, \sigma')\delta^{pq}~,
\end{equation}
that implies the following spectral representation
\begin{align}
\begin{split}\label{eq:Boneloop}
\nptf{J^p_\mu(\sigma) J^q_\rho(\sigma')}\vert_{\text{full, one-loop}} & = \int_{-\infty}^{+\infty} d \nu \, B^{pq}(\nu)_{\text{one-loop}} \,(\Omega_\nu^{(1)})_{\mu\rho}(\sigma,\sigma'
)~, \\
B^{pq}(\nu)_{\text{one-loop}}& = -\left(T-\frac{3}{4\pi}\right)\, \delta^{pq}~.
\end{split}
\end{align}
By setting again $d=1$ and using \eqref{eq:ourCtB} we obtain 
\begin{equation}
\left(C_t\right)_{\text{one-loop}} =  \frac{1}{\pi} \left(T-\frac{3}{4\pi}\right) = \frac{\sqrt{\lambda}}{2\pi^2}-\frac{3}{4\pi^2} + \mathcal{O}\left(\frac{1}{\sqrt{\lambda}}\right)\,,
\end{equation}
that indeed satisfies \eqref{eq:CtBrel} to subleading order.
 
Let us further comment on the particular choice of scheme. 
The continuation of $N_{\mathrm{y}}$ is closely related to the continuation of the number of scalar fields that is routinely implemented in perturbative calculations in $\mathcal N=4$ SYM in the supersymmetry-preserving DRED scheme (see e.g.~\cite{chakraborty2026} for recent results).\footnote{When dimensionally reducing to $D=4-2\eps$, four-dimensional vectors become $D$ dimensional vectors with extra $2\eps$ ``evanescent scalar'' components. The evanescent scalar can be equivalently kept inside 4 dimensional vector ($D=4$ vector and $N_{\text{scalars}}=6$) or separated ($D=4-2\eps$ vector and $N_{\text{scalars}}=6+2\eps$) \cite{Capper:1979ns}. Notice that in both formulations of boundary DRED, the evanescent scalar is there, so the continuation of $N_{\mathrm y}$ does not depend on which formulation is used.} In the holographic dual, this prescription corresponds to an AdS$_{5-2\eps}\times S^{5+2\eps}$ geometry. The line defect in $\mathcal{N}=4$ SYM is then dual to a string in this geometry.  
Assuming that the worldsheet dimension is continued by keeping fixed the transverse dimensions, we indeed get a $d+1=2-2\eps$-dimensional worldsheet QFT with $N_{\mathrm{y}}=5+2\eps$ massless scalars. It might be possible to derive this prescription also from purely two-dimensional considerations, for instance by studying how to preserve supersymmetry on the continued worldsheet. We leave this for future work.\footnote{Another aspect that requires further scrutiny is that, as already observed, the continuation with fixed $m_\mathrm{x}^2=2$ does not preserve the relation $\Delta_{\mathrm{x}}=d+1$ expected for the displacement operators. On the other hand since we are fixing $N_\mathrm{x}=3$ the interpretation of the fields $\mathrm{x}$ as transverse fluctuations appears preserved by the continuation. We note that anyway our final result for $C_\mathrm{t}$ is not affected by changing $N_\mathrm{x}$ and/or $m_\mathrm{x}^2$ at order $\mathcal{O}(\varepsilon')$, see \eqref{eq:resultbyfield}.}

\paragraph{Comparison with the naive approach} To further support the tilt definition via broken currents in AdS$_2$ as in \ref{eq:BOEJ}, as opposed to the naive holographic identification of $t^p$ with the massless modes $\mathrm{y}^p$, it is useful to see what happens if we evaluate the one-loop correction to the two-point function $\nptf{\mathrm{y}_p(\sigma) \mathrm{y}_q(\sigma')}$. Notice that in this case, only the tadpole diagrams contribute to the calculation.
We first focus on the $\mathrm{y}$ and $\theta$ sectors, where the relevant diagrams are given respectively by
\begin{align}
\begin{split}
    \begin{tikzpicture}[baseline=(current bounding box.center)]
    \label{diag_naive}
    \begin{feynman}
        \vertex [dot,label=below:\(\frac16 R_{turs}\del \mathrm{y}^t \cdot \del \mathrm{y}^r \mathrm{y}^u \mathrm{y}^s\)](a) {};
        \vertex [dot,label=above:\(\mathrm{y}_q\)](b) [right=of a]{};
        \vertex [dot,label=above:\( \mathrm{y}_p\)](c) [left=of a]{};
        \diagram*{
          (c) -- [scalar] (b),
          a --[scalar, out=135,in=45,loop,min distance=1.6cm] a,
        };
    \end{feynman}
    \end{tikzpicture} &=\frac13 R_{pq} G_d(\sigma,\sigma^\prime) G_d\big\vert_0\,,\\
\begin{tikzpicture}[baseline=(current bounding box.center)]
      \begin{feynman}
        \vertex [dot,label=below:\( \frac{1}{8} (\del \mathrm{y})^2 \bar{\theta}\Gamma_{123}\theta \)](a) {};
        \vertex [dot,label=above:\( \mathrm{y}_q\)](b) [right=of a]{};
        \vertex [dot,label=above:\(\mathrm{y_p}\)](c) [left=of a]{};
        \diagram* {
          (c) -- [scalar] (b),
          a --[out=135,in=45,loop,min distance=1.6cm] a,
        };
      \end{feynman}
    \end{tikzpicture}&= \delta_{pq} G_d(\sigma,\sigma^\prime)\frac{1}{4}N_\theta S_0\,.
    \end{split}
\end{align}
By using the results in \eqref{eq:propagators_dimreg}, we observe that in dimreg the above diagrams read
  \begin{align*}
      \mathrm y &: G_d(\sigma,\sigma^\prime)\delta_{pq} \frac13(N_{\mathrm{y}}-1)G_d\big\vert_0 = G_d(\sigma,\sigma^\prime)\delta_{pq}\frac13\left( \frac{1}{\pi\eps'}+\frac{1}{2\pi}\right)\,,\notag\\
       \mathrm \theta &: G_d(\sigma,\sigma^\prime)\delta_{pq} \frac{1}{4}N_\theta S_0= G_d(\sigma,\sigma^\prime)\delta_{pq}\left( -\frac{1}{\pi\eps'}+\frac{1}{\pi}\right)\,.
  \end{align*}
The sum of the two is clearly divergent. Moreover, also in this case, it is easy to see that the $\mathrm{x}$ sector contributes with finite diagrams only. We conclude that the full one-loop correction to the $\mathrm{y}$ two-point function is divergent in contrast with the result in \eqref{eq:JJoneloop}. Thus, the identification in \ref{eq:BOEJ} and the inclusion of the current diagrams are crucial ingredients to obtain a finite (and correct) result for $\left(C_t\right)_{\text{one-loop}}$. Equivalently, this means that the BOE coefficient $b_{\mathrm{y}t}$ is UV divergent at subleading order. The calculation in terms of bulk current allows to unambiguously determine what UV divergence is reabsorbed by $b_{\mathrm{y}t}$.

\section*{Acknowledgements}
We thank D.~Bonomi, T.~Canneti, S.~Giombi, S.C.~Lanza, P.~Niro, W.~M\"uck, R.~Roiban, D.~Seminara and A.~Tseytlin for discussions.
We thank the Isaac Newton Institute
for Mathematical Sciences, Cambridge (UK), for support and hospitality during the programme “Quantum Field Theories with Boundaries, Impurities and Defects” where work
on this paper was undertaken. This work was supported by EPSRC grant EP/Z000580/1. The work of FC and VF is supported by the DFG Research Unit 5582 “Modern foundations of Scattering Amplitudes", project number 508889767.  LDP acknowledges support by INFN Iniziativa Specifica ST\&FI. VF is supported by the DFG via the Heisenberg Professorship program 506208580. RS is supported by the DFG via the  Research Training Group  2575 ``Rethinking Quantum Field Theory''.

\appendix

\section{The dual string: \texorpdfstring{AdS$_2$}{AdS2} Lagrangian}
\label{Polyakov}

This appendix derives the Green–Schwarz fluctuation Lagrangian through quartic order around the \AdSd	
  classical worldsheet associated with the straight half-BPS Wilson line. Equivalent expressions have appeared previously in Refs.\cite{beccaria2026, Giombi:2017cqn}; here we obtain the result through a covariant Riemann-normal-coordinate (RNC) expansion.

\paragraph{Notation and conventions}
We use the following conventions, here and in the main text: Ten-dimensional target space coordinates are denoted by $x^m$ ($m = 0,1,\cdots,9$), and $e^{\underline{a}}_m$ denotes the corresponding vielbein, with $a=0,\dots 9$ a flat tangent-space index. Worldsheet coordinates are $\sigma=\sigma^\alpha$, $\alpha=0,1$, with corresponding zweibein  $\unie^{\flati{\alpha}}_\alpha$ ($\flati\alpha = 0,1$).
We denote the Levi-Civita symbol by $\epsilon^{\alpha\beta}$, with $\epsilon^{01}=1$, and the corresponding   tensor by $\varepsilon^{\alpha\beta}=\frac{\epsilon^{\alpha\beta}}{\sqrt{-g}}$. The $32\times32$ ten-dimensional gamma matrices $\Gamma^a$  satisfy the mostly-plus Clifford algebra
$\{\Gamma^a,\Gamma^b\}=2\eta^{ab}$,
while curved-index gamma matrices are defined through the vielbein, $\Gamma^m=e^m{}_a\Gamma^a$. In most formulas we work in an orthonormal frame and, when no ambiguity arises, suppress the distinction between curved and flat target-space indices for notational simplicity.

\paragraph{Green-Schwarz action in Polyakov form} 
We consider fluctuations of the Green–Schwarz (GS) type IIB superstring around a classical string configuration, eventually specializing to the \AdSd minimal surface in the AdS$_5\times S^5$ background dual to the straight Wilson line.
 In particular, in the Polyakov formulation, the GS action reads \cite{Cvetic:1999zs, Tseytlin:1996hs}
\begin{equation}
    \label{P:GS}
    S_P = -\frac1{4\pi \alpha'} \int \dd^2 \sigma \left( \sqrt{-h} h^{\alpha\beta} G_{\alpha\beta} - \varepsilon^{\alpha\beta} B_{\alpha\beta} \right)\,,
\end{equation}
where $h_{\alpha\beta}$ is the auxiliary metric. $G_{\alpha\beta}$ and $B_{\alpha\beta}$ found in \cite{Grisaru:1985fv, Wulff:2013kga} are given here to quadratic order in fermionic fields
\begin{align}
\label{P:G}
	G_{\alpha\beta} &= \gamma_{\alpha\beta} - 2i \bar{\theta}^I \Gamma_{(\alpha} \left(\mathcal{D}_{\beta)}\right)^{IJ} \theta^J\,,\\
\label{P:B}
	B_{\alpha\beta} &= \mathrm{B}_{\alpha\beta} - 2i \bar{\theta} ^I\sigma_3^{IJ}\Gamma_{[\alpha} \left(\mathcal{D}_{\beta]}\right)^{JK} \theta^K\,.
\end{align}
Since we are interested in one-loop corrections to correlators of bosonic fluctuations, we will not need higher orders in fermions.
In \eqref{P:G}, $\gamma_{\alpha\beta}$  is the induced worldsheet metric associated with the fluctuating string embedding (whereas $g_{\alpha\beta}$, introduced below, denotes its value on the classical solution, i.e. the \AdSd metric) and $\mathrm{B}_{\alpha\beta}$ is the pullback of the NSNS two-form. Here $\mathcal{D}_{\alpha}$ denotes the generalized covariant derivative acting on the type-IIB GS fermions. Its purely geometric part is the pullback of the ten-dimensional spinor covariant derivative $D_\alpha\equiv \partial_\alpha+
 \frac14\partial_\alpha x^n\,\omega_{npq}\Gamma^{pq}$,
 so that~\cite{Wulff:2013kga,Bigazzi:2024biz,Canneti:2025rsp,Martucci:2003gc}
\begin{equation}
\label{P:D}
	\left(\mathcal{D}_\alpha\right)^{IJ} = D_\alpha \delta^{IJ} +\frac18 H_{\alpha pq} \Gamma^{pq} \sigma_3^{IJ} +\frac18 S^{IJ} \Gamma_\alpha\,,
\end{equation}
where $H =\dd \mathrm{B}$, 
while the matrix $\mathcal S$ encodes the couplings to the Ramond–Ramond (RR) field strengths $F$ as \cite{Wulff:2013kga}
\begin{equation}    \label{P:SIIB}
	S = -e^{\phi} \left(i\sigma_2 F_{m} \Gamma^{m} +\sigma_1\frac1{3!} F_{mnp} \Gamma^{mnp} +\frac{i\sigma_2}{2\cdot 5!}F_{mnpqs} \Gamma^{mnpqs}\right)~.
\end{equation}
Above, $\phi$ is the dilaton and $\sigma_a$ ($a= 1,2,3$) are the Pauli matrices. In~\eqref{P:G} and~\eqref{P:B}, $\theta^I$ ($I=1,2$) is a doublet of ten-dimensional  Majorana-Weyl spinors of the same chirality. The Nambu-Goto formulation of the superstring action is obtained from the Polyakov one by setting the auxiliary metric in \eqref{P:GS} on-shell through the field equation
\begin{equation}
	\label{NG:h.eom}
	h^{\alpha\gamma} G_{\gamma\beta} - \frac12 \delta^\alpha_\beta h^{\gamma\delta} G_{\gamma\delta} =0\,.
\end{equation}
A semiclassical analysis for the GS action can be pursued by expanding the fields $G_{\alpha\beta}$ and $B_{\alpha\beta}$ in terms of fluctuations around a minimal string configuration and then truncating the action at the required order in the fluctuations. The interaction terms for the fluctuations can be written in a manifestly covariant way using Riemann normal coordinates (RNC) and geodesic expansion analysis to perform the expansion of the GS action~\cite{Alvarez-Gaume:1981exa,Atick:1986jr, Forini:2015mca, Bigazzi:2023oqm,Bigazzi:2024biz, Canneti:2025rsp,Singh:2023olv,Faraggi:2011ge}.  
 In what follows, we restrict to type-IIB supergravity backgrounds with vanishing NSNS  $\mathrm{B}$-field.

\paragraph{Auxiliary metric expansion} In the Polyakov formulation,  the worldsheet metric $h_{\alpha\beta}$ is an  independent auxiliary field. In a semiclassical expansion, it is decomposed into its classical value and a fluctuation as
\begin{equation}
    \label{temp:h}
    h^{\alpha\beta} = \bar h^{\alpha \beta} + \delta h^{\alpha\beta}\,,
\end{equation}
where $\bar h^{\alpha\beta}$ is determined by the Virasoro constraint in \eqref{NG:h.eom}.
Expanding the inverse metric and its determinant in powers of $\delta h^{\alpha\beta}$, the Weyl-invariant combination $\sqrt{-h}\,h^{\alpha\beta}$ appearing
in~\eqref{P:GS} takes, through fourth order, the form~\footnote{See e.g.\cite{Canneti:2025rsp}}
\begin{equation}
\label{P:unih.expand}
	\unih^{-1} + \delta\unih + \frac{1}{4}\unih^{-1}\tr(\delta\unih \unih)^2 - \frac{1}{6} \unih^{-1}\tr(\delta\unih \unih)^3 + \frac1{8}\unih^{-1}\left[\tr (\unih\delta \unih)^4-\frac34 \left(\tr(\delta\unih \unih)^2 \right)^2\right]+ \mathcal{O}(\delta \unih^5)\,,
\end{equation}
where we have defined the unimodular metric
\begin{equation}
\label{P:unih.def}
	\unih^{\alpha\beta} = \sqrt{-\bar h} \bar h^{\alpha\beta}\,,
\end{equation}
and its inverse, $\unih_{\alpha\beta}$. $\delta \unih^{\alpha\beta}$ denotes the traceless combination 
\begin{multline}
\label{P:delta.unih.def}
	\delta \unih = \sqrt{-\bar h}\bigg(1-\frac{1}{2}\tr\delta h \bar h+ \frac{1}{4}\tr(\delta h \bar h)^2 + \frac18(\tr \delta h \bar h)^2 - \frac16\tr(\bar h\delta h)^3 \\-\frac18
  \tr(\bar h\delta h)^2\tr \bar h\delta h-\frac{1}{48} (\tr\bar h\delta h)^3 \bigg) \left(\delta h -\frac{1}{2} \bar h^{-1}\tr \delta h \bar h\right)+ \mathcal{O}(\delta h^5)\,.
\end{multline}
As we stated before, the field equations for the auxiliary metric set the unimodular metric to satisfy\footnote{In what follows all the worldsheet indices are raised and lowered with the induced metric $g_{\alpha\beta}$.}
\begin{equation}
\label{P:unigh.background}
	\unih^{\alpha\beta} = \unig^{\alpha\beta}\equiv \sqrt{-g} g^{\alpha\beta}~,
\end{equation}
where $g_{\alpha\beta}$ is the classical (bosonic) induced metric on the worldsheet.

\paragraph{Bosonic fluctuation expansion} We expand the bosonic part of the action (\ref{P:GS}) around a classical string embedding $\bar{x}^m(\sigma)$  using a covariant background-field expansion based on the exponential map and Riemann normal coordinates (RNC)~\cite{Alvarez-Gaume:1981exa, Forini:2015mca, Bigazzi:2023oqm, Canneti:2025rsp}. The fluctuation $\xi^m(\sigma)$ is a target-space vector based at $\bar{x}^m(\sigma)$, and the full embedding is defined by $\xi: \bar{x} \to x = \exp_{\bar{x}}(\xi)$. This construction organizes the expansion covariantly in terms of geometric tensors evaluated on the reference embedding. In particular, the expansion of the pullback tangent vector takes the form
\begin{align}
\label{RNC:emb}
\partial_\alpha x^m & = \partial_\alpha \bar{x}^m+ D_\alpha \xi^m +\bigg[-\frac13 R^m{}_{plq} \xi^p \xi^q-\frac 1{12}  \left(D_n R^m{}_{plq} \right)\xi^n \xi^p \xi^q\\
      \notag
    	&\quad -\left(\frac 1{60}  D_s D_k R^m{}_{plq} +\frac{1}{45}R^m{}_{pqr}R^r{}_{skl}\right)\xi^s\xi^k \xi^p \xi^q\bigg]\partial_\alpha \bar{x}^l+ \cdots\,.
\end{align}
Here $D_\alpha \xi^m$ denotes the pullback of the target-space covariant derivative acting on $\xi^m$.  The components of $\xi$ are fields living on the worldsheet since they only depend on $\left(\sigma^0,\sigma^1\right)$. For notational simplicity, bars on background quantities will be suppressed below whenever no ambiguity can arise.  Up to quartic order in $\xi$, the induced metric changes as follows
\begin{align}
\label{NG:g.exp}
    \gamma_{\alpha\beta}= g_{\alpha\beta} + \mathcal{A}^{(1)}_{\alpha\beta} + \mathcal{B}^{(2)}_{\alpha\beta} +\mathcal{C}^{(3)}_{\alpha\beta} + \mathcal{D}^{(4)}_{\alpha\beta}\,, 
\end{align}
where 
\begin{align}
\notag
   \mathcal{A}^{(1)}_{\alpha\beta} &= - 
   \left(g_{mn}D_\alpha( \partial_\beta x^m)\xi^n + g_{mn}D_\beta( \partial_\alpha x^n)\xi^m \right)= -2 g_{mn} K_{(\alpha\beta)}^m \xi^n\,, \\
   \label{RNC:A_B_C_D}
    \mathcal{B}^{(2)}_{\alpha \beta}&=
    \left(g_{mn}D_\alpha \xi^m D_\beta \xi^n- R_{mpnq} \,\xi^p\xi^q\, \partial_\alpha x^m\partial_\beta x^n \right)\,,\\
    \notag
   \mathcal{C}^{(3)}_{\alpha \beta} &=- \frac{1}{3}\left(D_l R_{mpnq} \,\xi^l\xi^p\xi^q\, \partial_\alpha x^m\partial_\beta x^n- 4 R_{mpnq} \,\xi^n\xi^p\, D_{(\alpha}\xi^q \partial_{\beta)} x^m\right)\,,\\
   \notag
   \mathcal{D}^{(4)}_{\alpha \beta}&= \frac{1}{12}\bigg[6\, D_l R_{mpnq} \,\xi^l\xi^n\xi^p\, D_{(\alpha} \xi^q\partial_{\beta)} x^m+ 4 R_{mpnq} \,\xi^n\xi^p\,D_\alpha \xi^m D_\beta \xi^q
   \notag\\
   \notag
   &+ \left(-D_sD_l R_{mpnq} \,  +4 R^r{}_{slm}R_{rpqn}\right)\xi^s \xi^l\xi^p\xi^q\,  \partial_\alpha x^m\partial_\beta x^n\bigg]\,.
\end{align}
For a classical minimal surface, the extrinsic curvature $K_{\alpha\beta}^m$ has vanishing trace.  Moreover, several terms in~\eqref{RNC:A_B_C_D} drop out or simplify in the case of maximally symmetric manifolds (e.g. $D_l R_{mpnq} =0$). 
The bosonic Polyakov action then takes the form
\begin{equation}
\label{P:action.expand}
	S_{B} =T \int  \dd^2 \sigma\, \sqrt{-g} \left(\mathcal{L}^{(0)}_B + \mathcal{L}^{(1)}_B + \mathcal{L}^{(2)}_B + \mathcal{L}^{(3)}_B+ \mathcal{L}^{(4)}_B+\cdots \right)~, \quad T = \frac1{2\pi\alpha'}\,,
\end{equation}
with
\begin{align}
\label{P:Lbos}
\mathcal{L}^{(0)}_{B} &=-1\,,\notag\\
\mathcal{L}^{(1)}_{B} &=-\frac12 g^{\alpha\beta}\mathcal{A}^{(1)}_{\alpha\beta} \,,\notag\\
	\mathcal{L}^{(2)}_{B} &= -\frac12\left(g^{\alpha\beta}\mathcal{B}^{(2)}_{\alpha\beta} + \frac{1}{2}\Tr(\delta\unih \unig)^2 +  \frac{1}{\sqrt{-g}}\delta \unih^{\alpha\beta}\mathcal{A}^{(1)}_{\alpha\beta}\right)\,,\notag\\
   \mathcal{L}^{(3)}_{B} &= -\frac12\left(g^{\alpha\beta}\mathcal{C}^{(3)}_{\alpha\beta}-\frac{1}{3}\Tr(\delta\unih \unig)^3  + \frac{1}{\sqrt{-g}}\delta \unih^{\alpha\beta}\mathcal{B}^{(2)}_{\alpha\beta}\right)\,,\\
  \mathcal{L}^{(4)}_{B} &=-\frac12\left[g^{\alpha\beta}\mathcal{D}^{(4)}_{\alpha\beta}  + \frac{1}4 \Tr(\delta\unih \unig)^2  g^{\alpha\beta}\mathcal{B}^{(2)}_{\alpha\beta}+ \frac1{4}\left(\Tr (\unig\delta \unih)^4-\frac34 \left(\Tr(\delta\unih \unig)^2 \right)^2\right)+ \frac{1}{\sqrt{-g}}\delta \unih^{\alpha\beta}\mathcal{C}^{(3)}_{\alpha\beta}\right]\,.\notag
\end{align}
The equation of motion for the background embedding follows from the linear term and imposes  $g^{\alpha\beta}K^m_{\alpha\beta}=0$. 
To simplify the calculation, we focus on cases in which the extrinsic curvature of the worldsheet is identically vanishing, hence $\mathcal{A}^{(1)}_{\alpha\beta} =0$. 
This is true for the case of interest here, the minimal surface  in AdS$_5\times S^5$ corresponding to the straight Wilson line at the boundary.\footnote{Other worldsheet configurations characterized by $\mathcal{A}^{(1)}_{\alpha\beta} =0$ could be found by looking at strings lying on the tip of the cigars of confining models as in \cite{Canneti:2025rsp,Bigazzi:2024biz, Castellani:2024ial}. }

We next eliminate the auxiliary-metric fluctuation
 $\delta\unih^{\alpha\beta}$. 
Once $A^{(1)}_{\alpha\beta}=0$, it is decoupled from the embedding fluctuations at quadratic order, but the cubic Lagrangian contains the mixed term
$\delta\mathfrak h^{\alpha\beta}B^{(2)}_{\alpha\beta}$.
This term is removed by the  following traceless shift
  \cite{Forini:2015mca,Canneti:2025rsp}
\begin{equation}
    \label{P:delta.h.shift}
	\delta \unih^{\alpha\beta}  = \delta\unih^{\prime\,\alpha\beta} -\left( \unig^{\gamma(\alpha} \mathcal{B}^{(2)}_{\gamma}{}^{\beta)}- \frac12 \unig^{\alpha\beta}\mathcal{B}^{(2)}_{\gamma}{}^\gamma \right)~,
\end{equation}
after which we rename $\delta\unih^{\prime\,\alpha\beta}$ as $\delta \unih^{\alpha\beta}$. Up to terms beyond quartic order, the Lagrangian becomes
\begin{align}
\label{P:Lbos_shift}
	\mathcal{L}^{(2)}_{B} &= -\frac12\left(g^{\alpha\beta}\mathcal{B}^{(2)}_{\alpha\beta} + \frac{1}{2}\Tr(\delta\unih \unig)^2\right)\,,\notag\\
   \mathcal{L}^{(3)}_{B} &= -\frac12\left(g^{\alpha\beta}\mathcal{C}^{(3)}_{\alpha\beta}-\frac{1}{3}\Tr(\delta\unih \unig)^3\right)\,,\\
  \mathcal{L}^{(4)}_{B} &=-\frac12\bigg[g^{\alpha\beta}\mathcal{D}^{(4)}_{\alpha\beta}  - \frac12 \left(\mathcal{B}^{(2)\,\alpha\beta}\mathcal{B}^{(2)}_{\alpha\beta} - \frac12 \left(g^{\alpha\beta}\mathcal{B}^{(2)}_{\alpha\beta}\right)^2\right)+\frac{1}{(-g)}\delta \unih ^{\alpha\lambda}g_{\lambda\rho} \delta \unih^{\rho\sigma}\left(\mathcal{B}^{(2)}_{\sigma\alpha}- \frac14 g_{\sigma\alpha}\mathcal{B}^{(2)\,\lambda}_{\lambda}\right)\notag \\
  &+\frac{1}4 \Tr(\delta\unih \unig)^2  g^{\alpha\beta}\mathcal{B}^{(2)}_{\alpha\beta}+ \frac1{4}\left(\Tr (\unig\delta \unih)^4-\frac34 \left(\Tr(\delta\unih \unig)^2 \right)^2\right)+ \frac{1}{\sqrt{-g}}\delta \unih^{\alpha\beta}\mathcal{C}^{(3)}_{\alpha\beta}\bigg]\,.\notag
\end{align}
The remaining mixed term $\delta \unih^{\alpha\beta}\mathcal{C}^{(3)}_{\alpha\beta}$ in $\mathcal{L}^{(4)}_{B}$  is removed by an analogous higher-order shift. Eliminating the auxiliary fluctuation through its algebraic equation of motion then gives
\begin{align}
\label{P:Lagrangian_assumption_3}
	\mathcal{L}^{(2)}_{B} &= -\frac12 g^{\alpha\beta}\mathcal{B}^{(2)}_{\alpha\beta} \,,\notag\\
   \mathcal{L}^{(3)}_{B} &= -\frac12 g^{\alpha\beta}\mathcal{C}^{(3)}_{\alpha\beta}\,,\\
  \mathcal{L}^{(4)}_{B} &=-\frac12\bigg[g^{\alpha\beta}\mathcal{D}^{(4)}_{\alpha\beta}  - \frac12 \left(\mathcal{B}^{(2)\,\alpha\beta}\mathcal{B}^{(2)}_{\alpha\beta} - \frac12 \left(g^{\alpha\beta}\mathcal{B}^{(2)}_{\alpha\beta}\right)^2\right)\bigg]\,.\notag
\end{align}
The same result follows by expanding the Nambu–Goto action directly, providing an explicit check at the level of the semiclassical expansion of the equivalence between the Polyakov and Nambu–Goto formulations~\cite{FRADKIN1982413}. Upon specializing the general Riemann-normal-coordinate expansion~\eqref{P:Lagrangian_assumption_3} to the AdS$_5\times S^5$ background and to the straight-string solution, the resulting quartic bosonic Lagrangian is equivalent, up to integrations by parts, to the one obtained by expanding the static-gauge action in explicit target-space coordinates~\cite{beccaria2026}.

\paragraph{Fermionic sector of the fluctuation expansion} Working to quadratic order in the GS fermions, we now expand the fermionic sector in powers of the bosonic RNC fluctuation, still under the assumptions of vanishing bosonic NSNS two-form and vanishing extrinsic curvature. Here we focus on the type IIB case, the corresponding Type IIA construction is presented in Appendix~\ref{appendix:IIA}. The fermionic action is given by \cite{Cvetic:1999zs, Tseytlin:1996hs},
\begin{equation}
	\label{P:GS.action}
	S_F = \frac i{2\pi \alpha'} \int \dd^2 \sigma \left[\sqrt{-h} h^{\alpha\beta}\bar{\theta}^I \Gamma_{\alpha} \left(\mathcal{D}_{\beta}\right)^{IJ} \theta^J - \varepsilon^{\alpha\beta} \bar{\theta}^I \sigma_3^{IJ}\Gamma_{\alpha} \left(\mathcal{D}_{\beta}\right)^{JK}\theta^K\right]~.
\end{equation}
Using a RNC-based approach, we expand \eqref{P:GS.action} around the classical configuration. We define 
\begin{equation}
    \bar\theta^I \Gamma_{\alpha} \left(\mathcal{D}_{\beta}\right)^{IJ}\theta^J\bigg|_{x} = \bar\theta^I \Gamma_{\alpha} \left(\mathcal{D}_{\beta}\right)^{IJ}\theta^J\bigg|_{\bar x}+  \bar \theta^I  \left(\mathcal{O}^{(1)}_{\alpha\beta}\right)^{IJ} \theta^J + \bar \theta^I\left(\mathcal{O}^{(2)}_{\alpha\beta}\right)^{IJ}\theta^J\,.
    \label{eq:fluctuation_fer}
\end{equation}
The superscript $r=1,2$ counts  powers of the bosonic fluctuation $\xi$. Since the classical fermionic background vanishes, the bilinears $\bar\theta^I(\mathcal{O}^{(r)}_{\alpha\beta})^{IJ}\bar\theta^J$ have total fluctuation order $r+2$.  They are defined by the following expressions~\footnote{Equations  \eqref{O1} and \eqref{O2} follow from the RNC expansions of the pullback vielbein, spin connection and RR matrix~\cite{McArthur:1983fm}. 
See~\cite{Arutyunov:2015mqj} for a similar analysis.}
\begin{eqnarray}
\label{O1}
  && \!\!\!\!\! \bar \theta^I  \left(\mathcal{O}^{(1)}_{\alpha\beta}\right)^{IJ} \theta^J  = D_\alpha\xi^m\, \bar\theta^I\, \Gamma_m D_\beta \theta^I + 
  \frac{1}{8}\left(D_\alpha\xi^m\partial_\beta x^n +D_\beta\xi^n\partial_\alpha x^m\right)\, \bar\theta^I\,\Gamma_m S^{IJ}\Gamma_n  \theta^J\\\nonumber
&&~~~+ \frac{1}{8}\xi^p\bar\theta^I\,  \Gamma_\alpha \left(D_p S\right)^{IJ}\, \Gamma_\beta \theta^J\,,\\
\label{O2}
&&   \!\!\!\!\!      \bar \theta^I  \left(\mathcal{O}^{(2)}_{\alpha\beta}\right)^{IJ} \theta^J= 
        \frac{1}{8}D_\alpha\xi^m D_\beta \xi^n\bar\theta^I\,\Gamma_m S^{IJ}\Gamma_n  \theta^J\\\nonumber
        &&~~~+\frac{1}{8} \left(D_\alpha\xi^m\partial_\beta x^n +D_\beta\xi^n\partial_\alpha x^m\right)\xi^p\, \bar\theta^I\,\Gamma_m \left(D_pS\right)^{IJ}\Gamma_n  \theta^J \\\nonumber
        &&~~~
     - \frac{1}{16}\left[ e_l ^{\underline{a}}e_\beta ^{\underline{b}} R^l{}_{p\alpha q}+ e_l^{\underline{b}} e_\alpha ^{\underline{a}}R^l{}_{p\beta q}\right] \xi^p \xi^q  \bar \theta^I \Gamma_{\underline{a}} S^{IJ}\Gamma_{\underline{b}} \theta^J+ \frac{1}{16}\xi^p \xi^q\bar \theta^I \Gamma_\alpha\left( D_pD_qS \right)^{IJ}\Gamma_\beta\theta^J\\\nonumber
     &&~~~
    -\frac{1}{2}R^n{}_{p\alpha q}\xi^p \xi^q\,  \bar \theta^I\Gamma_n D_\beta\theta^I
  + \frac{1}{8}D_\beta \xi^n \xi^pR_{pn ql}\,  \bar \theta^I   \Gamma_\alpha \Gamma^{ql}\theta^I\,.
\end{eqnarray}
Above, a worldsheet index $\alpha$ appearing on a target-space tensor denotes pullback with the tangent vector $\partial_\alpha \bar{x}^m$.

Combining now the RNC expansion~\eqref{eq:fluctuation_fer} with (\ref{P:unih.expand}), the terms quadratic in the fermions through fourth order in the total fluctuations take the form~\footnote{Here the superscript denotes the total fluctuation order, $\mathcal{L}_{F}^{(2)}\sim\theta^{2}$,
 $\mathcal{L}_{F}^{(3)}\sim\xi\,\theta^{2}$, $
\mathcal{L}_{F}^{(4)}\sim\xi^{2}\theta^{2}$.
The pure $\theta^4$  sector is not included.}
\begin{eqnarray}
\label{P:action.expand.F}
	S_{F} &=&T \int  \dd^2 \sigma\, \sqrt{-g} \left(\mathcal{L}^{(2)}_F + \mathcal{L}^{(3)}_F+ \mathcal{L}^{(4)}_F+\cdots \right)\,,\\ 
\label{P:Lfer}
	\mathcal{L}^{(2)}_{F} &=& i\bar{\theta}^I\left[g^{\alpha\beta}\delta^{IJ}- \frac{1}{\sqrt{-g}}\varepsilon^{\alpha\beta}  \sigma_3^{IJ}\right]\Gamma_{\alpha} \left(\mathcal{D}_{\beta}\right)^{JK}\theta^K\,,\notag\\
    \mathcal{L}^{(3)}_{F}&=& i \bar{\theta}^I\left[g^{\alpha\beta}\delta^{IJ}- \frac{1}{\sqrt{-g}}\varepsilon^{\alpha\beta}  \sigma_3^{IJ}\right]\left(\mathcal{O}^{(1)}_{\alpha\beta}\right)^{JK}\theta^K  +i \frac{1}{\sqrt{-g}}\bar \theta^I\delta \unih^{\alpha \beta}\Gamma_{\alpha} \left(\mathcal{D}_{\beta}\right)^{IJ}\theta^J\,,\\
    \mathcal{L}^{(4)}_{F} &=&i \bar{\theta}^I\left[g^{\alpha\beta}\delta^{IJ}- \frac{1}{\sqrt{-g}}\varepsilon^{\alpha\beta}  \sigma_3^{IJ}\right]\left(\mathcal{O}^{(2)}_{\alpha\beta}\right)^{JK}\theta^K\notag\\
    &+&
    i \bar{\theta}^I\left(\frac{1}{\sqrt{-g}}\delta \unih^{\alpha \beta}\left(\mathcal{O}^{(1)}_{(\alpha\beta)}\right)^{IJ} +\frac{1}{4}g^{\alpha\beta}\Gamma_{\alpha} \left(\mathcal{D}_{\beta}\right)^{IJ}\Tr(\delta\unih \unig)^2\right)\theta^J\,.\notag
\end{eqnarray}
Applying the same auxiliary-metric shift used in the bosonic sector and subsequently eliminating the remaining auxiliary fluctuation through its algebraic equation of motion, we obtain
\begin{align}
\label{P:Lfer_4}
	\mathcal{L}^{(2)}_{F} &= i\bar{\theta}^I\left[g^{\alpha\beta}\delta^{IJ}- \frac{1}{\sqrt{-g}}\varepsilon^{\alpha\beta}  \sigma_3^{IJ}\right]\Gamma_{\alpha} \left(\mathcal{D}_{\beta}\right)^{JK}\theta^K\,,\notag\\
    \mathcal{L}^{(3)}_{F}&= i \bar{\theta}^I\left[g^{\alpha\beta}\delta^{IJ}- \frac{1}{\sqrt{-g}}\varepsilon^{\alpha\beta}  \sigma_3^{IJ}\right]\left(\mathcal{O}^{(1)}_{\alpha\beta}\right)^{JK}\theta^K \,,\\
    \mathcal{L}^{(4)}_{F} &=i \bar{\theta}^I\left[g^{\alpha\beta}\delta^{IJ}- \frac{1}{\sqrt{-g}}\varepsilon^{\alpha\beta}  \sigma_3^{IJ}\right]\left(\mathcal{O}^{(2)}_{\alpha\beta}\right)^{JK}\theta^K\notag\\
    &+  i\bar{\theta}^I\left(-\mathcal{B}^{(2)\,\alpha\beta}\Gamma_{\alpha} \left(\mathcal{D}_{\beta}\right)^{IJ}+ \frac1{2} \mathcal{B}^{(2)\,\gamma}_{\gamma} \Gamma^{\rho} \left(\mathcal{D}_{\rho}\right)^{IJ}\right)\theta^J\,.\notag
\end{align}
Equations~\eqref{P:Lagrangian_assumption_3} and~\eqref{P:Lfer_4}
give the covariant fluctuation action through fourth order in the
fluctuations and to quadratic order in the fermions, prior to fixing
worldsheet diffeomorphisms and $\kappa$-symmetry. The reduction to the
eight transverse bosonic fields and the sixteen physical fermionic
components appearing in the gauge-fixed action~\eqref{eq:AdS2Lagrangian}
is carried out below.

\paragraph{$\kappa$-symmetry gauge fixing} We now fix $\kappa$-symmetry and identify the physical fermionic degrees of freedom. Using the two-dimensional identity~\cite{Canneti:2025rsp}
\begin{align}
\label{P:gamma_relation}
\Gamma_\alpha \varepsilon^{\alpha\beta} =\unig^{\alpha\beta} \Gamma^{(3)}\Gamma_\alpha\,,\qquad \Gamma^{(3)} =  \frac12 \varepsilon^{\alpha \beta} \Gamma_{\alpha\beta}\,,
\end{align}
the tensor structure appearing in the quadratic fermionic
Lagrangian can be rewritten as 
\begin{equation}
\label{P:Proj}
   \frac12 \left[g^{\alpha\beta}\delta^{IJ}- \frac{1}{\sqrt{-g}}\varepsilon^{\alpha\beta}  \sigma_3^{IJ}\right]\Gamma_{\alpha} = \frac12 g^{\alpha\beta}\left[\delta^{IJ}-  \sigma_3^{IJ}\Gamma^{(3)}\right]\Gamma_{\alpha} = g^{\alpha\beta}\left(\Pi_-\delta^{I1}+\Pi_+\delta^{I2} \right)\Gamma_{\alpha}\,,
\end{equation}
where we introduced the projectors
\begin{equation}
    \Pi_\pm = \frac 12\left(1 \pm \Gamma^{(3)}\right)\,.
\end{equation}
By looking at the quadratic term in \eqref{P:Lfer_4}, a natural $\kappa$-symmetry gauge fixing is \cite{beccaria2026,Beccaria:2025xry, Kallosh:1997ky,Seibold:2024oyr}
\begin{align}
\label{KS_gauge_fixing}
     &\bar \theta_1 \Pi_+ =0\,,\quad \Pi_+\theta_1 = \theta_1\,,\notag\\
     &\bar \theta_2 \Pi_- =0\,,\quad \Pi_-\theta_2 = \theta_2\,.
\end{align}
For the $\mathrm{AdS}_2$ straight-string solution in
$\mathrm{AdS}_5\times S^5$, this gauge has the useful property that the cubic fermionic Lagrangian $\mathcal L_F^{(3)}$ vanishes 
\cite{beccaria2026}, as recalled explicitly below. In the main text, we use the following notation $\theta = \theta_1+\theta_2$.

\paragraph{Static gauge} 
We now decompose the ten bosonic fluctuations $\xi^m$ into components
tangent and normal to the classical worldsheet, 
\begin{equation}
\label{P:long_trans}
    \xi^ m = t^m_{\underline{\alpha}} \zeta^{\underline{\alpha}} + N^m_{\underline{i}} \chi^{\underline{i}}\,,\quad  t^m_{\underline{\alpha}} = \partial_\alpha x^m \unie^{\alpha}_{\underline{\alpha}} \,.
\end{equation}
Here, we introduced the normal vectors $ N^m_{\underline{i}}$, $(    \underline{i}=1,\cdots,8)$, which are chosen orthonormal and orthogonal to the tangent directions \footnote{The vectors $ N^m_{\underline{i}}$ are understood to span the normal bundle.
The latter displays an $SO(8)$ symmetry implementing the freedom to
rotate the normals. This symmetry is gauged on the worldsheet via the introduction of the gauge connection $A_{\flati{ij}\alpha}$ as $D_\alpha \chi^{\flati i} = \partial_\alpha \chi^{\flati i}-A^{\flati i}{}_{\flati{j}\alpha}\chi^{\flati j} $. Interestingly, the normal vectors
could have an obvious interpretation as part of the vielbein basis in an adapted frame to the worldsheet $N^m_{\flati i} = e^m_{\flati i}$.}
\begin{align}
  & N^m_{\flati i} t_{\flati \alpha}^n g_{mn } =0\,,\\
  & N^m_{\flati i}N^n_{\flati j}g_{mn} =\delta_{\flati i\flati j}\,,
\end{align}
and together with the tangent vectors satisfy the completeness relation
\begin{equation}
    \eta^{\flati{\alpha\beta}} t_{\flati \alpha}^m t_{\flati \beta}^n + \delta^{\flati {ij}}N^m_{\flati i}N^n_{\flati j}  = g^{mn}\,.
\end{equation}
At the level of the covariant derivative, this decomposition implies the following identities (see e.g.\cite{Forini:2015mca, Bigazzi:2023oqm,Canneti:2025rsp, Faraggi:2011ge})
\begin{align}
     &t^{\flati \alpha}_m D_\alpha \xi^m = D_\alpha \zeta^{\flati \alpha} -K^{\flati \alpha}_{m\alpha} N^m_{\flati i}\chi^{\flati i}\,,\notag\\
     & N_m^{\flati i} D_\alpha \xi^m  = D_{\alpha}\chi^{\flati i} + \zeta^{\flati \alpha} N_m^{\flati i} K^{m}_{\flati \alpha \alpha}\,.
\end{align}
Worldsheet diffeomorphism invariance allows us to set fluctuations $\zeta^{\flati \alpha}$ tangent to the worldsheet to zero. 
For the straight-string embedding considered here, both the
extrinsic curvature and the normal-bundle connection vanish, 
$K^m_{\alpha\beta}=0$, 
$A_{\flati{ij}\alpha}=0$. Together with static gauge, this reduces the covariant derivative
of the fluctuation to~\footnote{The string configurations falling in this class of embeddings do not restrict only to AdS$_2$ minimal worldsheets in “AdS$\times M$” backgrounds. Indeed, for instance, also winding string configurations considered in e.g. \cite{Bigazzi:2023oqm,Canneti:2025rsp, Bigazzi:2024biz} are characterized by vanishing $K^m_{\alpha \beta}$ and  $A_{\flati{ij}\alpha}$. }
\begin{equation}
     D_\alpha \xi^m = N^m_{\flati i} D_\alpha \chi^{\flati i}\,.
\end{equation}

\paragraph{Comments on the AdS$_2$ minimal string configuration} In specializing to the case of an AdS$_2$ minimal string configuration in AdS$_5\times S^5$, some helpful relations for maximally symmetric spacetimes are needed. In particular, in flat index notation 
\begin{align}
\label{HB:riemann_components}
    &R_{\underline{ijkl}} = -\left(\delta_{\underline{ik}}\delta_{\underline{jl}}-\delta_{\underline{il}}\delta_{\underline{jk}}\right)\,, \quad \underline{i},\underline{j},\underline{k},\underline{l}\in \text{AdS}_5\,,\notag\\
    &R_{\underline{pqrv}} = \left(\delta_{\underline{pr}}\delta_{\underline{qv}}-\delta_{\underline{pv}}\delta_{\underline{qr}}\right)\,, \quad \underline{p},\underline{q},\underline{r},\underline{v}\in S^5\,,\\
    &R_{\alpha \underline{j}\beta \underline{l}} = -g_{\alpha\beta}\delta_{\underline{jl}}\,, \quad \underline{j},\underline{l}\in \text{AdS}_5\,,\notag\\
    &R_{\alpha \underline{i}\underline{j}\underline{l}}=0 \,, \quad \forall \,\underline{i},\underline{j},\underline{l}\,.\notag
\end{align}
The self-dual RR five-form for the AdS$_5\times S^5$ background is given as
\begin{equation}
    F_5 = G_5 + \ast G_5 \,,\quad  G_5  = -4 \text{vol}_{\text{AdS}_5}\,.
\end{equation}

Moreover, by writing the Riemann tensor in terms of the metric and using metric compatibility, it is possible to observe that all the interactions involving covariant derivatives of the Riemann vanish in both AdS$_5$ and $S^5$ sectors.  As a consequence, the bosonic cubic Lagrangian $\mathcal{L}_B^{(3)}$ in \eqref{P:Lagrangian_assumption_3} vanishes identically.

As a final remark, let us stress that using the $\kappa$-symmetry gauge fixing in eqs. \eqref{KS_gauge_fixing} the fermionic cubic term $\mathcal{L}_F^{(3)}$ in \eqref{P:Lfer_4} can be set to zero \cite{beccaria2026}.\footnote{It is helpful to notice that following properties hold: $\left[\Pi_\pm\,,\Gamma_i\right] =0\,, \quad \Pi_\pm \Gamma_\alpha = \Gamma_\alpha\Pi_\mp\,, \quad  D_p S = 0\,$. }\footnote{This is a specific property of the $\kappa$-symmetry gauge fixing of \eqref{KS_gauge_fixing}. Indeed, in the alternative and widely used $\theta_1 = \theta_2$ gauge fixing $\mathcal{L}_F^{(3)}$ does not vanish, and other boson-fermion interactions appear in the theory.}

\section{Type IIA extension and the ABJM straight string}
\label{appendix:IIA}

This appendix extends the covariant fluctuation expansion of Appendix~\ref{Polyakov} to Type IIA superstring backgrounds and then specializes it to the $\mathrm{AdS}_2$ string dual to the half-BPS Wilson line in ABJM theory.
The bosonic quartic interactions relevant for holographic defect correlators in this background have been
considered previously in Refs.\cite{Bianchi:2020hsz,Bianchi:2026pee}; here we
include the corresponding terms quadratic in the fermionic
fluctuations.  This provides a starting point for testing, in the $\mathrm{AdS}_4\times\mathbb{CP}^3$
setting, the regularization prescription discussed in the main text. A more detailed analysis will be provided in~\cite{ABJM}. 

The quartic expansion of the GS action in (\ref{P:Lagrangian_assumption_3}) and \eqref{P:Lfer_4} can be easily extended to the type IIA superstring case. The bosonic is unchanged, while the
terms quadratic in the Type IIA fermion take the form
\begin{align}
\label{P:Lfer_IIA}
	\mathcal{L}^{(2)}_{F} &= i\bar{\theta}\left[g^{\alpha\beta}- \frac{1}{\sqrt{-g}}\varepsilon^{\alpha\beta} \Gamma^{(11)}\right]\Gamma_{\alpha} \mathcal{D}_{\beta}\theta\,,\notag\\
    \mathcal{L}^{(3)}_{F}&= i \bar{\theta}\left[g^{\alpha\beta}- \frac{1}{\sqrt{-g}}\varepsilon^{\alpha\beta}  \Gamma^{(11)}\right]\mathcal{O}^{(1)}_{\alpha\beta}\theta \,,\\
    \mathcal{L}^{(4)}_{F} &=i \bar{\theta}\left[g^{\alpha\beta}- \frac{1}{\sqrt{-g}}\varepsilon^{\alpha\beta}  \Gamma^{(11)}\right]\mathcal{O}^{(2)}_{\alpha\beta}\theta+  i\bar{\theta}\left(-\mathcal{B}^{(2)\,\alpha\beta}\Gamma_{\alpha} \mathcal{D}_{\beta}+ \frac1{2} \mathcal{B}^{(2)\,\gamma}_{\gamma} \Gamma^{\rho} \mathcal{D}_{\rho}\right)\theta\,.\notag
\end{align}
Here $\mathcal{D}_{\alpha}$ denotes the generalized covariant derivative
acting on the Type IIA GS fermion, namely (see e.g.\cite{Wulff:2013kga, Canneti:2025rsp})
\begin{equation}
\label{NG:DIIA}
	\mathcal{D}_\alpha = \partial_\alpha +\partial_\alpha x
^n\left[\frac{1}{4}\omega_{npq}\Gamma^{pq} +\frac18 H_{npq} \Gamma^{pq} \geleven +\frac18 S \Gamma_n\right]~,
\end{equation}
with 
\begin{equation}
\label{NG:SIIA}
	S = e^{\phi} \left( \frac12 F_{mn} \Gamma^{mn} \geleven +\frac1{4!} F_{mnpq} \Gamma^{mnpq} \right)~. 
\end{equation}
In this case, $\theta$ is a 32-component  Majorana spinor. The $\mathcal{O}^{(i)}_{\alpha\beta}$ are defined as in \eqref{O1} and \eqref{O2} upon sending $S^{IJ} \to S$. Using the identity (\ref{P:gamma_relation}), the quadratic term in (\ref{P:Lfer_IIA}) can be written as
\begin{equation}
\label{L2Pi}
   \mathcal{L}^{(2)}_{F} = 2 i \bar\theta\Pi_-\unig^{\alpha\beta}\Gamma_{\alpha}\mathcal{D}_\beta\theta\,,
\end{equation}
where 
\begin{equation}
    \Pi_\pm = \frac 12\left(1 \pm \Gamma^{(11)}\Gamma^{(3)}\right)\,.
\end{equation}
We then fix $\kappa$-symmetry by imposing
\begin{equation}
\label{KS_gauge_fixingIIA}
    \Pi_-\theta= \theta\,.
\end{equation}

\paragraph{ABJM half-BPS Wilson line } The half-BPS Wilson line in ABJM theory is described holographically
by a fundamental Type IIA string whose classical worldsheet is an
$\mathrm{AdS}_2$ minimal surface in
$\mathrm{AdS}_4\times\mathbb{CP}^3$. The corresponding supergravity
background is~\cite{Aharony:2008ug}
\begin{align}
\label{ABJM_solution}
    &\mathrm{d}s^2_{10} = \frac{R^3}{4k}\left(\mathrm{d} s^2_{AdS_4} + 4\mathrm{d}s^2_{\mathbb C P^3}\right)\,,\notag\\
    & F_4 =\frac38 R^3\, \text{vol}_{AdS_4}\,,\quad F_2 = 2kJ \,,\quad e^{2\phi}= \frac{R^3}{k^3}\,,
\end{align}
where $J$ is the K\"ahler two form for $\mathbb CP^3$. The relation of the parameters entering in (\ref{ABJM_solution}) and the field theory ones is
\begin{equation}
    \frac{R^3}{4k} = \left(\pi\sqrt{2\lambda}\right)\,\qquad \lambda = \frac{N}{k}\,,
\end{equation}
where $\lambda$ is the 't Hooft coupling.
For later convenience, we rescale the background fields according to~\cite{Muck:2016hda}
\begin{equation}
    \mathrm{d}\hat s^2_{10} = \beta^2 \mathrm{d}s^2_{10}\,,\quad \hat \phi = 0\,,\quad \hat C_p = e^\phi C_p \beta^{\frac{p}{2}}\,\quad \beta = \frac{4k}{R^3}\,.
\end{equation}
For notational simplicity, we suppress the hats below. 
In these conventions, the background becomes~\footnote{Notice that the string tension is now $\hat T = T\beta^{-1}$. In this case, we have that $\hat T = \frac{1}{2\pi\alpha^\prime} \frac{R^3}{4k} = \frac{\sqrt{2\lambda}}{2}$. }
\begin{align}
\label{ABJM_solution_2}
    &\mathrm{d}s^2_{10} = \left(\mathrm{d} s^2_{AdS_4} + L^2\mathrm{d}s^2_{\mathbb C P^3}\right)\,,\notag\\
    & F_4 =3 \, \text{vol}_{AdS_4}\,,\quad F_2 = L^2 J \,,\quad e^{2\phi}= 1\,\quad L=2\,.
\end{align}
For a unit-radius $\mathbb{CP}^3$, we use the conventions
%
\begin{eqnarray}
  &&  \!\!\!\!\!\! \mathrm{d}s^2_{\mathbb C P^3}=\frac14\bigg[\mathrm{d}\alpha^2 +  \cos^2\frac\alpha 2 \left(\mathrm{d}\theta_1^2 + \sin^2\theta_1\mathrm{d}\varphi_1^2\right)+ \sin^2\frac\alpha 2 \left(\mathrm{d}\theta_2^2 + \sin^2\theta_2\mathrm{d}\varphi_2^2\right)\notag\\
   & &\qquad+ \cos^2 \frac\alpha 2\sin^2 \frac\alpha 2\left(\mathrm{d}\psi + \cos \theta_1 \mathrm{d}\varphi_1 - \cos \theta_2 \mathrm{d}\varphi_2\right)^2\bigg]\,,\\
&& \!\!\!\!\!\!   J = \frac12 \mathrm{d}A\,,\quad A = \frac12\left[\frac12\left(\cos^2\frac\alpha 2 -\sin^2\frac{\alpha}{2} \right)\mathrm{d}\psi +\cos^2\frac\alpha 2\cos\theta_1\mathrm{d}\varphi_1+ \sin^2\frac\alpha 2\cos\theta_2\mathrm{d}\varphi_2\right]\,.
\end{eqnarray}
The classical straight-string embedding is
\begin{equation}
\label{EmbeddingABJM}
    x^0 = t = \sigma^0,\qquad z = \sigma^1,
\end{equation}
with all remaining target-space coordinates held fixed. The induced
worldsheet metric is that of $\mathrm{AdS}_2$, and the embedding has vanishing extrinsic curvature. 

Imposing static gauge as in Appendix~\ref{Polyakov}(\ref{P:long_trans}), the eight transverse
bosonic fluctuations split into two modes $x^i$, $i=1,2$, normal to
the $\mathrm{AdS}_2$ surface inside $\mathrm{AdS}_4$, and six modes
$y^p$, $p=4,\ldots,9$, tangent to $\mathbb{CP}^3$. In a worldsheet-adapted orthonormal frame, the normal vectors along
$\mathbb{CP}^3$ may be identified with the inverse of the vielbein along the transverse directions. Notice that for the $\mathbb{C}P^3$ directions we will use the following conventions $N^m_{\flati p} = \frac 1L \hat e^m_{\flati p}$, where $\hat e^m_{\flati p}$ are the vielbein for $\mathbb{C}P^3$ with radius one. In what follows, we will drop the underlined indices for simplicity.

With the $\kappa$-symmetry gauge~\eqref{KS_gauge_fixingIIA}, the cubic fermionic Lagrangian 
$\mathcal{L}_F^{(3)}$ also vanishes for this background.

\paragraph{Quartic fluctuation Lagrangian on AdS$_2$}
The resulting gauge-fixed fluctuation Lagrangian, through quartic
order in the total fluctuations and keeping terms up to quadratic
order in the fermions, is
\begin{align}
\mathcal{L}_B^{(2)}
&=
-\frac12\left(
g^{\alpha\beta}\partial_\alpha\mathrm{x}^i\partial_\beta \mathrm{x}_i
+g^{\alpha\beta}\partial_\alpha \mathrm{y}^p\partial_\beta \mathrm{y}_p
+2\mathrm{x}^i \mathrm{x}_i
\right)\,,
\nonumber\\
\mathcal{L}_B^{(4)}
&=
-\frac12 \mathrm{x}^4
-\frac14 g^{\alpha\beta}\mathrm{x}^i \mathrm{x}_i\,
\partial_\alpha \mathrm{x}^i \partial_\beta \mathrm{x}_i
+\frac1{6L^2}\hat R_{pqks}g^{\alpha\beta}
\partial_\alpha \mathrm{y}^p
\partial_\beta \mathrm{y}^k
\mathrm{y}^q\mathrm{y}^s
\nonumber\\
&\qquad
+\frac18
\left(
g^{\mu\alpha}g^{\nu\beta}
+g^{\mu\beta}g^{\alpha\nu}
-g^{\mu\nu}g^{\alpha\beta}
\right)
\nonumber\\
&\qquad\quad\times
\Bigl(
\partial_\mu \mathrm{x}^i\partial_\nu \mathrm{x}_i\,
\partial_\alpha\mathrm{x}^j\partial_\beta \mathrm{x}_j
+2\partial_\mu \mathrm{x}^i\partial_\nu \mathrm{x}_i\,
\partial_\alpha \mathrm{y}^p\partial_\beta \mathrm{y}_p
\nonumber\\
&\hspace{5.0cm}
+\partial_\mu \mathrm{y}^p\partial_\nu \mathrm{y}_p\,
\partial_\alpha \mathrm{y}^q\partial_\beta \mathrm{y}_q
\Bigr)\,,
\displaybreak[4]\\
\mathcal{L}_F^{(2)}
&=
\frac{i}{2}\bar{\theta}\slashed{\mathcal{D}}\theta\,,
\nonumber\\
\mathcal{L}_F^{(4)}
&=
\frac{i}{8}
\Bigl[
\bigl(
g^{\alpha\beta}g^{\rho\mu}
-g^{\alpha\rho}g^{\beta\mu}
-g^{\alpha\mu}g^{\rho\beta}
\bigr)
\bigl(
\partial_\rho \mathrm{x}^i\partial_\mu\mathrm{x}_i
+\partial_\rho \mathrm{y}^p\partial_\mu\mathrm{y}_p
\bigr)
\nonumber\\
&\hspace{3.2cm}
+2g^{\alpha\beta}
\bigl(\mathrm{x}^i\mathrm{x}_i\bigr)
\Bigr]
\bar{\theta}\Gamma_\alpha\mathcal{D}_\beta\theta
\nonumber\\
&\qquad
+\frac{i}{4}\bar{\theta}
\Bigl[
g^{\alpha\beta}
\Bigl(
\partial_\alpha\mathrm{x}^i\partial_\beta\mathrm{x}^j
\frac18\Gamma_i S\Gamma_j
+\partial_\alpha \mathrm{y}^p\partial_\beta \mathrm{y}^q
\frac18\Gamma_p S\Gamma_q
\Bigr)
\nonumber\\
&\hspace{3.4cm}
+\bigl(\mathrm{x}^i\mathrm{x}_i\bigr)
\frac18\Gamma_\alpha S\Gamma_\beta
\Bigr]\theta
\nonumber\\
&\qquad
+\frac{i}{8}g^{\alpha\beta}\bar{\theta}
\Bigl(
-\mathrm{x}^i\partial_\alpha \mathrm{x}^j
\Gamma_\beta\Gamma_{ij}
+\frac12\Gamma_\alpha
\mathrm{y}^p\partial_\beta \mathrm{y}^q
\hat R_{pqmn}\Gamma^{mn}
\Bigr)\theta
\nonumber\\
&\qquad
-\frac{i}{4}\bar{\theta}\epsilon^{\alpha\beta}\Gamma^{(11)}
\Bigl[
\partial_\alpha\mathrm{x}^i\partial_\beta \mathrm{x}^j
\frac18\Gamma_iS\Gamma_j
+\partial_\alpha \mathrm{y}^p\partial_\beta \mathrm{y}^q
\frac18\Gamma_pS\Gamma_q
\Bigr]\theta\,.
\label{eq:AdS2LagrangianABJM}
\end{align}
Above, we have rescaled the fermion as
$\theta\to\frac12\theta$. The unit-radius
$\mathbb{CP}^3$ Riemann tensor is
\begin{equation}
\label{RiemannCP3}
    \hat R_{ijks} =  \left[\left(\delta_{ik}\delta_{js}-\delta_{is}\delta_{jk}\right)+2 J_{ij } J_{ks}+J_{ik } J_{js}-J_{i s} J_{jk }\right]\,, \quad i,j,k,s\in \mathbb{C}P^3\,,
\end{equation}
and $L=2$.\footnote{See \cite{Bianchi:2020hsz,Bianchi:2026pee} for a similar result for the quartic bosonic Lagrangian.}

The two transverse $\mathrm{AdS}_4$ modes have
$m_x^2=2$ ($\Delta=2$), whereas the six $\mathbb{CP}^3$ modes are
massless ($\Delta=1$). 
In \eqref{eq:AdS2LagrangianABJM}, we defined
\begin{align}
     \mathcal{D}_\alpha &= \Bigl[D_\alpha - \frac{1}{d+1}\Gamma_\alpha\mathcal{M}_F\Bigr]\,,\notag\\
    D_\alpha &= \Bigl(\partial_\alpha + \frac14 \omega_{\alpha mn} \Gamma^{mn} \Bigr)\,,\quad
    \mathcal{M}_F = \frac14\left(\frac12 L^2 J_{mn} \Gamma^{mn}\geleven + 3 \Gamma^{0123}\right)\,.
\end{align}
%


\section{Variation of the Lagrangian and Noether current}
\label{app:variation_noether}

In this appendix, we verify the invariance of the
Lagrangian~(\ref{eq:AdS2Lagrangian}) under the transformations~(\ref{eq:coset_variation}), and derive the associated Noether current.

At leading order in the transformation parameter $\varepsilon$,
the variations of the bosonic terms read
\begin{align}
\label{app:bos_variation}
\delta \mathcal{L}_B^{(2)}
&= \frac13 g^{\mu\nu} \partial_\mu\mathrm{y}^q\partial_\nu\mathrm{y}^p \varepsilon^m \mathrm{y}^n R_{mpqn}+ \cdots\,,\notag\\
\delta \mathcal{L}_B^{(4)} &=  \frac13 g^{\mu\nu} \partial_\mu\mathrm{y}^q\partial_\nu\mathrm{y}^p \varepsilon^m \mathrm{y}^n R_{mpnq}+ \cdots\,.
\end{align}
For the fermionic sector, neglecting terms beyond the fluctuation
order retained in the action, one finds
\begin{align}
\label{app:ferm_variation}
\delta \mathcal{L}_F^{(2)}
&=  -\frac{i}{16}\eps^p \partial_\nu \mathrm{y}^q\,g^{\mu\nu}R_{pqmn}\bar{\theta}\Gamma_\mu\Gamma^{mn}\theta+ \cdots\,,\notag\\
\delta \mathcal{L}_F^{(4)} &= \frac{i}{16}\eps^p \partial_\mu \mathrm{y}^q\,R_{pqmn}\bar{\theta}\Gamma^\mu\Gamma^{mn}\theta + \cdots\,.
\end{align}
Here $\delta\mathcal L_F^{(2)}$ arises from the fermionic
transformation, whereas $\delta\mathcal L_F^{(4)}$ receives the
relevant contribution from the bosonic variation, since
$\delta\theta=\mathcal O(y\theta)$.   Notice that in deriving $\delta \mathcal{L}_F^{(4)}$ we used the fact that $\left[\slashed{\mathcal{D}}, \Gamma^{ab}\right] =0$. Combining~\eqref{app:bos_variation} and \eqref{app:ferm_variation} the variations cancel,
confirming invariance of the action to the order considered.

We next derive the associated Noether current and its off-shell
identities. Consider an action of the form
\begin{equation}
S = \int \dd^2 \sigma \sqrt{g}\,\mathcal{L}[\mathrm{y},\bar{\theta},\theta]\,,
\end{equation}
then, a symmetry of $S$ is such that the corresponding Lagrangian variation satisfies 
\begin{equation}
  \delta \mathcal{L} =  D_\mu H^\mu\,.
  \label{eq:defsym}
\end{equation}
In particular, by defining $\delta\mathcal{L} = \varepsilon^p\delta_p \mathcal{L}$, we have that
\begin{equation}
\delta_p \mathcal{L}
= \frac{\partial\mathcal{L}}{\partial \mathrm{y}^q}\delta_p \mathrm{y}^q
+ \frac{\partial\mathcal{L}}{\partial(\partial_\mu \mathrm{y}^q)}D_\mu\delta_p \mathrm{y}^q
+ \frac{\partial\mathcal{L}}{\partial\theta}\delta_p\theta
+ \frac{\partial\mathcal{L}}{\partial(D_\mu\theta)}D_\mu\delta_p\theta + \delta_p\bar{\theta} \frac{\partial\mathcal{L}}{\partial\bar{\theta}}
+ \delta_p\bar{\theta}\overleftarrow{D}_\mu \frac{\partial\mathcal{L}}{\partial(\bar{\theta}\overleftarrow{D}_\mu)}\,.
\end{equation}
Integrating by part and using the fact  that the present Lagrangian contains
no dependence on~$\bar{\theta}\overleftarrow{D}_\mu$, we obtain
\begin{equation}
-D_\mu\Bigl(
\frac{\partial\mathcal{L}}{\partial(\partial_\mu \mathrm{y}^q)}\delta_p \mathrm{y}^q
+ \frac{\partial\mathcal{L}}{\partial(D_\mu\theta)}\delta_p\theta
- H^\mu
\Bigr)
= -D_\mu J^\mu_p
= E_{\mathrm{y}^q}\delta_p \mathrm{y}^q + E_\theta \delta_p\theta + \delta_p\bar{\theta} E_{\bar{\theta}}\,.
\label{eq:Noether_formal}
\end{equation}
where $E_{y^q}$, $E_\theta$ and $E_{\bar\theta}$ denote the corresponding Euler--Lagrange expressions.
We use the standard covariant integration-by-parts identity
$\partial_\mu(\sqrt g\,V^\mu)=\sqrt g\,D_\mu V^\mu$. For the transformations considered here the Lagrangian is invariant
without the addition of a total derivative, so that $H^\mu=0$. By using the product rule in the left-hand-side, the $D_\mu(\partial\mathcal{L}/\partial(\cdots))\,\delta_p(\cdots)$ terms cancel and one is left with
\begin{equation}
-\frac{\partial\mathcal{L}}{\partial(\partial_\mu \mathrm{y}^q)}D_\mu\delta_p \mathrm{y}^q
-\frac{\partial\mathcal{L}}{\partial(D_\mu\theta)}D_\mu\delta_p\theta
=
\frac{\partial\mathcal{L}}{\partial \mathrm{y}^q}\delta_p \mathrm{y}^q
+
\frac{\partial\mathcal{L}}{\partial\theta}\delta_p\theta
+
\delta_p\bar{\theta} \frac{\partial\mathcal{L}}{\partial\bar{\theta}}\,.
\end{equation}
For the present Lagrangian, the individual terms entering this
identity are
\begin{align}
  \frac{\del\mathcal{L}}{\del(\del_\mu \mathrm{y}^q)}D_\mu\delta_p \mathrm{y}^q &=  \frac13 g^{\mu\nu} \partial_\nu\mathrm{y}^q\partial_\mu\mathrm{y}^n \mathrm{y}^m R_{qmpn}\,,\notag\\
  \frac{\del\mathcal{L}}{\del(D_\mu\theta)}D_\mu\delta_p \theta &= -\frac{i}{16}R_{pqmn}\bar{\theta}\left( \slashed{\del} \mathrm{y}^q \Gamma^{mn}\theta + \mathrm{y}^q \Gamma^{mn}\slashed{D}\theta\right)\,, \notag\\
  \frac{\del\mathcal{L}}{\del \mathrm{y}^q}\delta_p \mathrm{y}^q &=  \frac13 g^{\mu\nu} \partial_\nu\mathrm{y}^q\partial_\mu\mathrm{y}^n \mathrm{y}^m R_{mqpn} + \frac{i}{16}R_{pqmn}\bar{\theta} \slashed{\del} \mathrm{y}^q \Gamma^{mn}\theta\,,\\
  \delta_p \bar{\theta} \frac{\del\mathcal{L}}{\del\bar{\theta}} + \frac{\del\mathcal{L}}{\del\theta}\delta_p \theta &= \frac{i}{16}R_{pqmn}\bar{\theta} \mathrm{y}^q \Gamma^{mn}\slashed{D}\theta\,.\notag
\end{align}
\section{Example of diagram evaluation}
\label{app:example}
In this appendix we provide details of two representative one-loop
diagram evaluations used in Section~\ref{Section:Bremsstrahlung}.

\paragraph{$\mathrm{y}$-loop tadpole diagram}


\begin{equation}
  \begin{tikzpicture}[baseline=(current bounding box.center)]
      \begin{feynman}
        \vertex [dot,label=below:\(\frac16 R_{turs}\del \mathrm{y}^t \cdot \del \mathrm{y}^r \mathrm{y}^u \mathrm{y}^s\)](a) {};
        \vertex [dot,label=above:\(-\del'_\nu \mathrm{y}_q\)](b) [right=of a]{};
        \vertex [dot,label=above:\(-\del_\mu \mathrm{y}_p\)](c) [left=of a]{};
        \diagram*{
          (c) -- [scalar] (b),
          a --[scalar, out=135,in=45,loop,min distance=1.6cm] a,
        };
      \end{feynman}
    \end{tikzpicture} 
    =
\frac16 R_{turs} \int_{\text{EAdS}_2} \dd\Sigma\, g^{\alpha\beta} \nptf{\partial_\mu \mathrm{y}_p(\sigma)\,
\mathrm{y}^u \mathrm{y}^s \partial_\alpha \mathrm{y}^t \partial_\beta\mathrm{y}^r(\Sigma)
\partial_\nu \mathrm{y}_q(\sigma')}\,.
\end{equation}
There are four distinct Wick contractions, namely
\begin{multline}
\label{eq:E_tadpole}
      2 \times  \frac16 R_{turs}  \int_{\Sigma}  g^{\alpha\beta} \Bigg( \nptf{\wick{\partial_\mu \c2 {\mathrm y}_p(\sigma)\, \c2 {\mathrm y}^u \c3 {\mathrm y}^s \partial_\alpha \c1 {\mathrm y}^t \partial_\beta \c1 {\mathrm y}^r(\Sigma) \partial_\nu \c3 {\mathrm y}_q(\sigma')}}  
+ \nptf{\wick{\partial_\mu \c1 {\mathrm y}_p(\sigma)\, \c1 {\mathrm y}^u \c2 {\mathrm y}^s \partial_\alpha \c2 {\mathrm y}^t \partial_\beta \c3 {\mathrm y}^r(\Sigma) \partial_\nu \c3 {\mathrm y}_q(\sigma')}} \\
+\nptf{\wick{\partial_\mu \c2 {\mathrm y}_p(\sigma)\, \c3 {\mathrm y}^u \c1 {\mathrm y}^s \partial_\alpha \c1 {\mathrm y}^t \partial_\beta \c2 {\mathrm y}^r(\Sigma) \partial_\nu \c3 {\mathrm y}_q(\sigma')}}
+ \nptf{\wick{\partial_\mu \c3 {\mathrm y}_p(\sigma)\, \c1 {\mathrm y}^u \c1 {\mathrm y}^s \partial_\alpha \c3 {\mathrm y}^t \partial_\beta \c2 {\mathrm y}^r(\Sigma) \partial_\nu \c2 {\mathrm y}_q(\sigma')}} \Bigg) \\
= \frac13 R_{tprq} \delta^{tr} \del_\alpha\del'_\beta G_d\big\vert_0 \int_{\Sigma}\del_\mu G_d(\sigma,\Sigma)\del'_\nu G_d(\sigma',\Sigma)   + \frac23 R_{tpqs}\delta^{st} \del^\alpha G_d\big\vert_0 \int_{\Sigma} \del^\Sigma_\alpha(\del_\mu G_d(\sigma,\Sigma)\del'_\nu G_d(\sigma',\Sigma))\\
+ \frac13  R_{puqs}\delta^{us} G_d\big\vert_0\,
\int_{\Sigma} g^{\alpha\beta}\del^\Sigma_\alpha\del_\mu G_d(\sigma,\Sigma)\del^\Sigma_\beta \del'_\nu G_d(\sigma',\Sigma)\,,
\end{multline}
The first two terms in \eqref{eq:E_tadpole} vanish, either because of the EOM identities (\ref{eq:EOMidentity_scalar}) (combined with setting $\delta^{d+1}(\sigma,\sigma)$ to zero in dimensional regularization) or because the first derivative of the scalar propagator
vanishes at coincident points. The remaining convolution can be evaluated using the
identity~\eqref{eq:convolution}. As a result, the diagram contribution to the one-loop current two-point function is 
\begin{equation}
  \begin{tikzpicture}[baseline=(current bounding box.center)]
      \begin{feynman}
        \vertex [dot,label=below:\(\frac16 R_{turs}\del \mathrm{y}^t \cdot \del \mathrm{y}^r \mathrm{y}^u \mathrm{y}^s\)](a) {};
        \vertex [dot,label=above:\(-\del'_\nu \mathrm{y}_q\)](b) [right=of a]{};
        \vertex [dot,label=above:\(-\del_\mu \mathrm{y}_p\)](c) [left=of a]{};
        \diagram*{
          (c) -- [scalar] (b),
          a --[scalar, out=135,in=45,loop,min distance=1.6cm] a,
        };
      \end{feynman}
    \end{tikzpicture} 
    =
\frac13 R_{pq}  G_d\big\vert_0\partial_\mu\partial^\prime_\nu G_d(\sigma,\sigma^\prime)\,.
\end{equation}
\paragraph{$\theta$-loop current diagram}
\begin{align*}
  \begin{tikzpicture}[baseline=(current bounding box.center)]
      \begin{feynman}
        \vertex [dot,label=below:\( \frac{1}{4} \del_\mu \mathrm{y}_p \bar{\theta}\Gamma_{123}\theta \)](a) {};
        \vertex [dot,label=above:\(-\del'_\nu \mathrm{y}_q\)](b) [right=of a]{};
        \diagram* {
          (a) -- [scalar] (b),
          a --[out=135,in=45,loop,min distance=1.5cm] a,
        };
      \end{feynman}
    \end{tikzpicture}
&=
-\frac{1}{4}
\left\langle
\partial_\mu \mathrm{y}_p\,\bar\theta\,\Gamma_{123}\theta(\sigma)\,
\partial_\nu \mathrm{y}_q(\sigma')
\right\rangle
=
\frac{1}{4} \partial_\mu \partial'_\nu G_d(\sigma,\sigma^\prime)\,
\delta_{pq}\,\mathrm{tr}\,\Gamma_{123}\,S\big\vert_0 \\
&=
-\frac{1}{4} N_\theta\,\partial_\mu \partial'_\nu G_d(\sigma,\sigma^\prime)\,
\delta_{pq}\,S_0\,.
\end{align*}
The minus sign in the first line comes from fermionic anticommutation.  The second line follows from the coincident-point
propagator~\eqref{S0}--\eqref{S00} and
$\Gamma_{123}^2=-1$.

\bibliographystyle{utphys}
\bibliography{WL_and_bubble_Draft,GS_expansion,ABJM}

\end{document}